\documentclass[11pt, sort&compress]{iopart}

\usepackage[utf8]{inputenc}
\usepackage[english]{babel}
\usepackage[T1]{fontenc}
\usepackage{natbib}
\usepackage[usenames,dvipsnames]{color}
\usepackage{color}
\usepackage{xcolor}
\usepackage{mathtools}
\usepackage{graphicx}
\expandafter\let\csname equation*\endcsname\relax
\expandafter\let\csname endequation*\endcsname\relax
\usepackage{amsmath}
\usepackage{amsfonts}
\usepackage{braket}
\usepackage{soul}
\usepackage{booktabs}
\usepackage{comment}
\usepackage{subcaption}
\usepackage{float}
\usepackage[colorlinks=true, hyperfootnotes=true, breaklinks=true, citecolor=blue]{hyperref}
\usepackage{silence}
\expandafter\let\csname equation*\endcsname\relax
\expandafter\let\csname endequation*\endcsname\relax

\begin{document}

\title {Effect of Weak Measurement Reversal on Quantum Correlations in a Correlated Amplitude Damping 
Channel, with a Neural Network Perspective}
\markboth{Effect of Weak Measurement Reversal on Quantum Correlations in a CAD channel}{}

\author{Venkat Abhignan}
 \address{Department of Physics, National Institute of Technology, Tiruchirapalli - 620015, India.}
\address{Qdit Labs Pvt. Ltd., Bengaluru - 560092, India.}

\author{Bidyut Bikash Boruah}
\address{Raman Research Institute, Sadashivanagar, Bengaluru - 560080, India.}

\author{R. Srikanth}
\address{Poornaprajna Institute Of Scientific Research, Bengaluru - 562110, India.}

 \author{Ashutosh Singh}
\address{Department of Physics and Astronomy and Institute of Quantum Science and Technology, University of Calgary, Calgary, AB, T2N 1N4, Canada.}
\ead{Corresponding Author: asinghrri@gmail.com}
\date{\today} 
\begin{abstract}

We study the evolution of quantum correlations in Bell, Werner, and maximally entangled mixed states of two qubits subjected to correlated amplitude-damping channels. Our primary focus is to evaluate the robustness of entanglement as a resource for quantum information protocols such as dense coding, teleportation, and Einstein–Podolsky–Rosen (EPR) steering under the influence of noise. In addition, we investigate the behaviour of other quantum correlations, including quantum discord and coherence, and analyze their hierarchy under decoherence. To counteract the detrimental effects of the channels, we apply the weak measurement and quantum measurement reversal (WMR) protocol, comparing the effectiveness of single-qubit and two-qubit WMR techniques. Our results show that the two-qubit WMR protocol significantly outperforms the single-qubit approach in preserving quantum correlations. Furthermore, we employ a neural network model to enhance our analysis of the relationship between different quantum correlation measures during the evolution. Using a MATLAB-based artificial neural network with 80 neurons across three hidden layers and trained with the Levenberg–Marquardt algorithm, we successfully predict trace distance discord from other correlations, achieving low prediction errors. Besides, our analysis of the neural network weights suggests that concurrence and EPR steering have the most positive influence on the accurate discord predictions.


\end{abstract}
\maketitle

\section{Introduction}
\label{Introduction}

\textcolor{black}{Quantum coherence \cite{QCoh_Alexander17} is a fundamental property of quantum mechanics that enables a single quantum system to exist in a superposition of distinct physical states. It is a core feature of quantum systems and is central to many quantum phenomena. Quantum entanglement \cite{QCoh_Alexander17, Ent_Horodecki2009}, a specific form of nonclassical correlation, arises due to shared coherence between two or more subsystems of a composite quantum system. Although entanglement is deeply rooted in the superposition principle, it is essential to note that a quantum system can possess coherence without exhibiting any entanglement. Entanglement plays an indispensable role in many quantum information processing protocols \cite{Book-QIP_chuang}, such as quantum communication, cryptography, quantum computation, dense coding, teleportation, etc.}

Quantum decoherence \cite{Decoherence_Zurek03}, resulting from the inevitable interaction between a quantum system's degrees of freedom and its environment, leads to the loss of coherence and the degradation of entanglement. In some cases, decoherence causes entanglement to vanish in a finite time, a phenomenon known as entanglement sudden death (ESD) \cite{ESD_Yu04, ESD_Yu2009}. Since entanglement is a crucial resource for many quantum information applications, addressing decoherence's effects is paramount. One way to mitigate decoherence is via entanglement distillation protocols \cite{Distillation_Kwiat01}, which recover maximally entangled states from a set of partially decohered ones. However, these methods cannot restore entanglement once the system becomes entirely separable. Other approaches to control decoherence include decoherence-free subspaces \cite{DFS_Lidar98, DFS_Kwiat2000}, \textcolor{black}{dynamical decoupling \cite{DD_Viola1999, DD_Du2009, DD_Yan2022, DD_Singh2018}}, the quantum Zeno effect \cite{QZeno_Facchi04, QZeno_Maniscalco2008}, the weak measurement and quantum measurement reversal (WMR) protocol \cite{Kim:09, QMR_Korotkov10, QMR_Lee11, WMQMR_Kim11}, and the application of local unitary operations \cite{ESD_Rau2008, ESD_Singh17, ESD_Singh22, ESD_Behera2025} during the system's evolution under noise. However, decoherence-free subspaces require specific symmetries in the interaction Hamiltonian, which may not always be present.

The concept of reversing a weak quantum measurement was first proposed in the context of quantum error correction \cite{WMR_Koashi99}. Later, the WMR protocol was shown to suppress decoherence \cite{QMR_Korotkov10, QMR_Lee11} induced by an amplitude-damping channel (ADC). \textcolor{black}{Subsequently, the protocol was successfully used to protect entanglement \cite{Kim:09, WMQMR_Kim11, PhysRevLett.97.166805, Heibati_2022} under ADC. The WMR approach has also been demonstrated to enhance teleportation Fidelity \cite{PRAMANIK2013} and improve secret key rates for specific quantum key distribution protocols \cite{DATTA2017, Haseli_2019} under ADC.} The WMR protocol leverages the weak measurement process to combat decoherence actively and can prevent even the sudden death of entanglement.

\textcolor{black}{Recently, the weak measurement has also opened up new avenues for exploring the foundational aspects of quantum mechanics. As suggested by Einstein–Podolsky–Rosen \cite{EPR1935}, the nature of reality for non-commuting observables was tested on polarization-entangled photon pairs using joint weak measurements followed by post-selection \cite{calderon-losada-2020}. A weak measurement scheme without post-selection has been employed in Bell inequality tests to measure the Bell parameter for each polarization-entangled photon pair without destroying their entanglement \cite{EntPreserv_Virzi2024}. Also, two weak measurements on both photons of each polarization-entangled photon pair have been used to test the relativistic bound \cite{Nonlocality_Atzori2024} experimentally. Notably, these measurements are forbidden in the standard projective measurement scheme of quantum mechanics.}

In this work, we consider three types of entangled two-qubit states—the Bell state, the Werner state, and a maximally entangled mixed state (MEMS) \cite{MEMS1_Munro2001, MEMS2_Munro2001}—and study their evolution under a correlated amplitude damping (CAD) channel \cite{TCADC_Yeo2003}. We analyze the performance of the WMR protocol in protecting various quantum correlations. 
It is essential to clarify that in the context of quantum information theory, "memory" or "correlation" in the noise may refer to two different concepts: (i) the divisibility of the quantum dynamical map \cite{kumar2018non}, and (ii) temporal or spatial correlations in the environmental noise. Here, we focus on the latter notion—memory in the noise. In one of our earlier works \cite{ESD_NAASDK23}, we investigated the entanglement and coherence of MEMS under correlated noise and found that memory effects help preserve quantum features. In the present study, we extend this analysis to the Bell-state, Werner-state, and MEMS under a CAD channel and explore the evolution and protection of several other quantum correlations using the WMR protocol and their relevance to quantum information protocols. 

To counter the impact of decoherence, we employ the WMR protocol on one or both qubits and study its effectiveness in preserving quantum correlations. We compare the impact of the WMR protocol on various measures of correlation, including Concurrence ($C$) \cite{EntCADWMQMR_Xing16}, Jensen-Shannon divergence (JSD) \cite{PhysRevLett.116.150504}, trace distance discord (TDD) \cite{Ciccarello_2014}, EPR steering (QS) \cite{schneeloch}, dense coding capacity ($\chi$) \cite{DensCodCADWMQMR_Li19}, and teleportation Fidelity ($\mathcal{F}$) \cite{TeleportCADWMQMR_Li18, DensTeleport_Wang23}. Specifically, we explore and contrast the effectiveness of single- and two-qubit WMR protocols, which have not been thoroughly investigated in prior works. The motivation behind studying multiple correlation measures is their ability to probe different regimes and aspects of quantumness. The correlation measures above follow a hierarchy \cite{Hierarchy_Adesso2016}, with quantum coherence (JSD) and quantum discord (TDD) being the weakest characterization of quantumness, followed by quantum entanglement (quantified by Concurrence $C$), with quantum steering (QS) being the strongest among these. The evolution of quantumness in the system can thus be studied by examining the hierarchical behaviour of these correlations as they evolve under the influence of a CAD channel, and their protection under the WMR protocol. Two other quantities considered here that are also measures of quantum correlation are dense coding capacity ($\chi$) and quantum teleportation Fidelity ($\mathcal{F}$), which quantify the classical and quantum communication capacity, respectively.

\textcolor{black}{Quantum discord captures correlations beyond entanglement and shows non-monotonic behaviour with entanglement. Moreover, it isn't easy to compute, even with the complete information of the state available, e.g., by quantum state tomography. The optimization involved in its definition is the main source of this difficulty \cite{Huang_2014}. Machine learning methods have been increasingly used to address challenges in quantum state tomography \cite{Torlai2018, Schmale2022}, correlation detection \cite{Yosefpor_2020, Chen_2022, ROIK2022, Asif2023, Urena2024, Huang2025}, and correlation classification \cite{Sanavio2023, PhysRevA.107.032421}. Notably, machine learning has enabled the identification of effective entanglement witnesses without requiring full tomography \cite{Yosefpor_2020, PhysRevApplied.19.034058}, and has been used to detect and quantify various correlation measures, including quantum discord \cite{Liu2019, Taghadomi2025}. Motivated by these studies, we explore whether relatively accessible correlations --- such as coherence, entanglement, and steering --- can be used to predict the more elusive quantum discord (TDD), without presuming prior knowledge of the functional form of its evolution. Using machine learning, we aim to capture the nonlinear relationships between these correlations, thereby reducing the computational cost of directly calculating quantum discord and uncovering deeper insights into the structure of quantum correlations.}

To quantify coherence, a function must satisfy specific criteria to be considered a valid coherence measure \cite{baumgratz2014quantifying}. Several such measures exist \cite{PhysRevLett.116.150502, Adesso_2016}, generally classified into geometric and entropic categories. If a function satisfies the triangle inequality and the axioms of distance, it is considered a \textcolor{black}{distance-based} measure \cite{baumgratz2014quantifying}. Based on the quantum Jensen-Shannon divergence (JSD), an important measure was introduced in Ref. \cite{PhysRevLett.116.150504}. This measure, which exhibits both geometric and entropic characteristics, is defined as:
\begin{align}
\text{JSD}(\rho) =  \sqrt{S \left( \frac{\rho + \rho_{d}}{2} \right) - \frac{S(\rho)}{2} - \frac{S(\rho_{d})}{2}},
\end{align}
where $S(\rho)$ is the von Neumann entropy, and $\rho_{d}$ is the closest incoherent state (diagonal in the reference basis). For a given density matrix $\rho$, the von Neumann entropy is:
\begin{align}
S(\rho) = - \sum_i \lambda_i \log_2 \lambda_i,
\end{align}
where $\lambda_i$ are the eigenvalues of $\rho$.

To quantify entanglement, we use Concurrence \cite{Conc_Hill1997, Conc_Wooters1998}, a computable measure for pure and mixed two-qubit states. For a two-qubit density matrix $\rho$, Concurrence is defined as:
\begin{align}
C(\rho) = \max\{0, \sqrt{\lambda_1} - \sqrt{\lambda_2} - \sqrt{\lambda_3} - \sqrt{\lambda_4} \},
\end{align}
where the $\lambda_i$'s are the eigenvalues of the matrix $\rho \tilde{\rho}$ in decreasing order, and $\tilde{\rho} = (\sigma_y \otimes \sigma_y)\rho^*(\sigma_y \otimes \sigma_y)$ is the spin-flipped density matrix.

The remainder of the paper is structured as follows: Section~(\ref{Preliminaries}) introduces the framework of open quantum systems, the dynamics under correlated ADC, and the WMR protocol. It also presents the definitions of dense coding, quantum teleportation, EPR steering, and trace distance discord. Section~(\ref{Results}) discusses our key findings regarding the aforementioned quantum correlations under the CAD channel and the impact of one- and two-qubit WMR protocols. Section~(\ref{ANN}) describes the neural network model used to learn and predict trace distance discord from other correlation measures. Finally, Section~(\ref{summary}) summarizes our conclusions.

\section{Preliminaries} \label{Preliminaries}

\subsection{Correlated amplitude damping channel}
 
Let us consider a two-level quantum system with the lower and upper levels denoted as $\ket{0}_S$ and $\ket{1}_S$, respectively. If the environment is initially in the $\ket{0}_E$ state, then the amplitude damping (AD) channel leads to a coupling between the system and the environment that can be expressed \cite{Preskill_LectureNotes} by the following map:
\begin{align}
\begin{aligned}
    \ket{0}_S\ket{0}_E &\rightarrow \ket{0}_S\ket{0}_E,\\
    \ket{1}_S\ket{0}_E &\rightarrow \sqrt{\bar{p}}\ket{1}_S\ket{0}_E + \sqrt{p}\ket{0}_S\ket{1}_E,
\end{aligned}
\end{align}
where $0\leq p \leq 1$ is the decoherence strength of the AD channel with $\bar{p}=1-p$. The Kraus operators for the AD channel are obtained by tracing out the environment's degrees of freedom and giving
\begin{align}
    E_0 = \begin{bmatrix}
        1 &0\\
        0 &\sqrt{\bar{p}}
    \end{bmatrix},~
    E_1 = \begin{bmatrix}
        0 &\sqrt{p}\\
        0 &0
    \end{bmatrix}.
\end{align}

For a two-qubit memoryless AD channel, the evolution of an initial state $\rho$ can be expressed as
\begin{align}
    \rho_{AD} = \sum_{i,j=0}^1 E_{ij}\rho E_{ij}^\dagger~,
\end{align}
where $E_{ij} = E_i\otimes E_j$, $i,j=\{0,1\}$. For a memory or CAD channel, such a tensorial decomposition is not possible, and the evolution under a partially CAD channel \cite{TCADC_Yeo2003} with memory parameter $\eta$ can be described as 
\begin{align}
    \rho_{CAD} = (1 - \eta)\sum_{i,j=0}^1 E_{ij}\rho E_{ij}^\dagger + \eta \sum_{k=0}^{1}A_k\rho A_k^\dagger~.
    \label{Eq_CAD}
\end{align}

The memoryless AD channel can be recovered by setting $\eta = 0$, and for $\eta = 1$, one obtains a perfect CAD channel. The Kraus operators $A_k$ \cite{TCADC_Yeo2003} have the following form 
\begin{align}
    A_0 = \begin{bmatrix}
        1 &0 &0 &0\\
        0 &1 &0 &0\\
        0 &0 &1 &0\\
        0 &0 &0 &\sqrt{\bar{p}}
    \end{bmatrix},~~
    A_1 = \begin{bmatrix}
        0 &0 &0 &\sqrt{p}\\
        0 &0 &0 &0\\
        0 &0 &0 &0\\
        0 &0 &0 &0
    \end{bmatrix}~.
\end{align}

\textcolor{black}{The noise described by a CAD channel incorporates memory effects stemming from a common or structured environment. Such correlations arise naturally when a bath or channel is shared among subsystems, such as trapped ions, superconducting qubits, etc. Thus, the CAD channel characterizes experimentally relevant decoherence dynamics outside the Markovian limit.}

\subsection{Weak-measurement and quantum measurement reversal}

 The technique of weak measurement (WM) or partial collapse measurement \cite{Kim:09, QMR_Korotkov10, QMR_Lee11, WMQMR_Kim11} allows for the extraction of some information from a quantum system without collapsing it to one of its eigenstates. Therefore, the system's initial state can be recovered with a certain probability by applying a reversal operation. To protect the entanglement of the qubits from decoherence, before they pass through the CAD channel, they are subject to WM, which can be expressed as a non-unitary quantum operation. For the case of WM on both qubits, the WM operation assumes the form
 \begin{align}
     \mathcal{M}_{WM} = \begin{bmatrix}
         1 &0\\
         0 &\sqrt{1-{q}_1}
     \end{bmatrix}\otimes
     \begin{bmatrix}
         1 &0\\
         0 &\sqrt{1-{q}_2}
     \end{bmatrix}~,
 \end{align}
where $q_1$ and $q_2$ are the WM strengths for the first and second qubits, respectively. After the state evolves through the CAD channel, a quantum measurement reversal (QMR) operation is performed on both qubits. Similar to the WM operation, the QMR operation is a non-unitary one, which is given as
\begin{align}
    \mathcal{M}_{QMR} = \begin{bmatrix}
        \sqrt{1-{r}_1} &0\\
        0 &1
    \end{bmatrix}\otimes
    \begin{bmatrix}
        \sqrt{1-{r}_2} &0\\
        0 &1
    \end{bmatrix}~,
\end{align}
where $r_1$ and $r_2$ are the QMR strengths for the first and second qubits, respectively. For simplicity, we consider the case where the strength of the WM and QMR operations on both the qubits is the same, i.e., $q_1 = q_2 = q$ and $r_1 = r_2  = r$, when WMR is performed on both the qubits. The final evolved state after the sequence of prior WM, CAD, and QMR operations is given as
\begin{align}
    \textcolor{black}{\rho_\text{WMR}[\rho]} =  \mathcal{M}_{QMR} \mathcal{E}_{CAD}(\mathcal{M}_{WM}\rho\mathcal{M}_{WM}^\dagger)   \mathcal{M}_{QMR}^\dagger~.
\end{align}
\textcolor{black}{The optimal value of the QMR strength $r$ can be found numerically by maximizing the entanglement (Concurrence calculated in Appendix A, B, C.) of $\rho_\text{WMR}[\rho]$ for a given value of $p, q$.}  In the case of WM performed on only one of the qubits, say the second qubit, the WM and corresponding QMR operations are reduced to
\begin{align}
    \mathcal{M}_{WM} = \begin{bmatrix}
         1 & 0\\
         0 & 1
     \end{bmatrix}\otimes
     \begin{bmatrix}
         1 & 0\\
         0 & \sqrt{\bar{q}}
     \end{bmatrix},~\text{and}~~
         \mathcal{M}_{QMR} = \begin{bmatrix}
        1 & 0\\
        0 & 1
    \end{bmatrix}\otimes
    \begin{bmatrix}
        \sqrt{\bar{r}} & 0 \\
        0 & 1 
    \end{bmatrix},~
\end{align}
with $\bar{q}=1-q$ and $\bar{r}=1-r$.

\subsection{Capacity of Dense Coding}

Dense coding \cite{DenseCoding_Bennet92}, also known as superdense coding,  is a quantum communication protocol that allows two parties to communicate two bits of classical information by transmitting only one qubit using pre-shared entanglement between them. As a prerequisite of the protocol, Alice and Bob share an entangled pair, say $\rho$. The ideal dense coding protocol assumes a noiseless quantum channel for entanglement distribution between Alice and Bob. 
Alice applies the Pauli operator $U_x$ with probability $P_x$ on her qubit and sends it to Bob through a noiseless quantum channel. Then Bob performs a joint measurement on both the qubits to retrieve the classical information. Given that channels are inevitably noisy in all realistic scenarios, the noise effects must be included in the dense coding protocol. Here, we consider the impact of the CAD channel on the entangled pair distribution from Charlie to Alice and Bob; hence, the WMR protocol is applied in this step, resulting in $\rho_\text{WMR}$. Subsequently, Alice applies the local unitary transformation on her qubit and sends it to Bob through a noiseless quantum channel for joint Bell-state measurement to recover the classical information.

Using the computational basis $\{\ket{0},\ket{1}\}$, the operators for quantum dense coding \cite{Book-QIP_chuang} can be expressed as
\begin{align}
\begin{aligned}
    & U_{00}\ket{x}  = \ket{x},
   && U_{01}\ket{x} = \ket{x+1(\text{mod}~2)},\\
    & U_{10}\ket{x} = e^{i\pi x}\ket{x},
   && U_{11}\ket{x} = e^{i\pi x}\ket{x+1\text{(mod}~2)}.
\end{aligned}
\end{align}

The evolved density matrix $\rho_\text{WMR}^*$ after dense coding on the distributed entanglement $\rho_\text{WMR}$ is given by
\begin{align}
    \rho_\text{WMR}^* = \frac{1}{4}\sum_{i=0}^3(U_i\otimes I)\rho_\text{WMR}(U_i^\dagger \otimes I)~,
\end{align}
where in the summation, we have replaced $\{00,01,10,11\}$ with $\{0,1,2,3\}$. A measure of the success of dense coding is the capacity of dense coding ($\chi$) \cite{DenseCoding_Hiroshima2001}, which is given by the Holevo quantity 
\begin{align}
    \chi = S(\rho_\text{WMR}^*) - S(\rho_\text{WMR})~,
\end{align}
where $S(\rho)$ is the von Neumann entropy of $\rho$. \textcolor{black}{This expression quantifies the extra amount of classical information that can be transmitted by exploiting entanglement. For example, $\chi > 1$, the protocol transmits more than one classical bit per qubit sent, demonstrating a clear quantum advantage over classical channels. In the ideal case of perfect dense coding, which is typically achieved with a maximally entangled state, $\chi$ reaches its maximum value of 2, meaning two bits of classical information are transmitted per qubit.}

\subsection{Fidelity of Quantum Teleportation}

 Quantum Teleportation \cite{Teleportation_Bennett93, Teleportation_Bouwmeester97} is a quantum communication protocol that, in a sense, is the opposite of superdense coding. Here, classical communication transfers an unknown quantum state without physically sending a qubit from the sender to the receiver, using pre-shared entanglement between them. Suppose Alice and Bob share an entangled state $\rho$ distributed through a noisy channel. Therefore, the WMR protocol is implemented to protect the entanglement of the initial state $\rho$, resulting in $\rho_\text{WMR}$. 
 
 It is essential to quantify the significance of an entangled state $\rho$ for teleportation. For this purpose, the fully entangled fraction (FEF) \cite{FEF_Bennett1996, FEF_Grondalski2002} provides a simple way to define the optimal Fidelity. The FEF for a quantum state $\rho$ in a $2\otimes 2$ Hilbert space is defined as the maximum overlap between the state $\rho$ and the maximally entangled pure states $\ket{\Phi}$, maximized for all $\ket{\Phi}$:
\begin{align}
    F(\rho) = \max_{\ket{\Phi}} \{\bra{\Phi}\rho\ket{\Phi}\}~,
\end{align}

where $F(\rho)$ is the FEF of $\rho$. It measures the closeness of the state $\rho$ to maximally entangled states. The optimum Fidelity of teleportation \cite{FEF_Grondalski2002} is related to FEF as
\begin{align}
    \mathcal{F}_{T_{2\otimes 2}}^{\rm max} = \frac{1}{3}\left[1+2F(\rho)\right].
\end{align}

This simple closed-form expression of FEF for a two-qubit system
was also given in Ref. \cite{FEF_Bennett1996}. Furthermore, it also provides a recipe to calculate the FEF:  let us consider an arbitrary density matrix $\rho$ of a pair of qubits. First, $\rho$ needs to be written in the basis $\{\ket{e_j}\}$ defined as
\begin{align}\label{fef_basis}
 \begin{aligned}
     \ket{e_1} =  \ket{\Phi^+},~
     \ket{e_2} =  i\ket{\Phi^-},~
     \ket{e_3} =  i\ket{\Psi^+},~
     \ket{e_4} =  \ket{\Psi^-},
 \end{aligned}
 \end{align}
 where $|\phi^{\pm}\rangle$ and $|\psi^{\pm} \rangle$ are the Bell-states \cite{Book-QIP_chuang}. A real vector represents a fully entangled state written on this basis. Therefore, to find the fully entangled fraction of the state, we need to maximize $\bra{e} \rho \ket{e}$ over all the real vectors $\ket{e_j}$. This maximum value is given by the largest eigenvalue of the real part of $\rho$. Therefore, when an entangled state $\rho$ is written in the basis defined in Eq.~(\ref{fef_basis}), FEF is simply the largest eigenvalue of the real part of $\rho$.

\subsection{Trace Distance Discord}

To determine suitable methods for computing quantum correlations beyond entanglement, geometric quantum discord based on the Hilbert-Schmidt norm was presented for mixed states \cite{GD2010}. However, typically for two-qubit systems, modified geometrical quantifier trace distance discord (TDD) \cite{Ciccarello_2014} is employed as a measure of such correlations because of the contractive nature \cite{contractive} and local ancilla problem \cite{ancilla} in geometric quantum discord. TDD is defined from the trace norm (Schatten 1-norm), the distance between the state of concern and the nearest zero discord states \cite{tdd2012,tddpra2013}.

According to local measurements on subsystem $a$, trace distance discord is the trace norm distance between state $\rho$ and the set of classical-quantum density matrices that show zero quantum discord \cite{Ciccarello_2014}. The definition of it is \begin{align}
\mathcal{T}(\rho ) =~^{\text{min}}_{\Pi ^{a}}\| \rho - \Pi ^{a}(\rho )\|, 
\end{align} where minimization over all potential depolarizing channels $\Pi^a$ are related to projective measurements in orthonormal basis from \begin{align} \label{defpi}
 \Pi^a(\cdot\cdot\cdot) = P_a \cdot\!\cdot\!\cdot P_a + Q_a \cdot\!\cdot\!\cdot Q_a\;,
 \end{align} with $P_a\equiv |\Phi\rangle_a\langle\Phi|$ defined from pure state $|\Phi\rangle$ and $Q_a = \mathbb{I}_a - P_a$ are rank-one projectors defined as its orthogonal complement. \textcolor{black}{This construction makes it clear that the TDD measures how much the state \(\rho\) is disturbed by the best possible local measurement on subsystem \(a\); a zero value indicates that \(\rho\) is already classical-quantum (i.e. it exhibits no quantum discord), while a nonzero value quantifies the presence of TDD}. Further, the Fano-Bloch decomposition on the $\rho$ is necessary for the analytical evaluation of TDD. Utilizing it, TDD can be expressed in compact form for the X-state $\rho$ as \begin{align}
    \mathcal{T}(\rho)=\frac{1}{2}\sqrt{\frac{\gamma_1^2 \max\{\gamma_3^2,\gamma_2^2+x_{A3}^2\}-\gamma_2^2 \min\{\gamma_3^2,\gamma_1^2\}}{\max\{\gamma_3^2,\gamma_2^2+x_{A3}^2\}- \min\{\gamma_3^2,\gamma_1^2\}+\gamma_1^2-\gamma_2^2}},
\end{align}
where, \begin{align}
    \gamma_1 = 2(\rho_{32}+\rho_{41}),\,\gamma_2 = 2(\rho_{32}-\rho_{41}),\,\gamma_3 = 1-2(\rho_{22}+\rho_{33}) \,\,\,\hbox{and}\,\,\, x_{A3} = 2(\rho_{11}+\rho_{22})-1.
\end{align}
The robustness of TDD was checked against Concurrence previously by us in a two-qubit spin-squeezing model under intrinsic decoherence \cite{Abhignan_2021}.

\subsection{EPR steering}

EPR steering is an alternative type of quantum correlation quantifier for nonlocality \cite{EPR1935, PhysRevLett.98.140402} intermediate between nonseparability \cite{PhysRevA.40.4277} and Bell nonlocality \cite{PhysRevLett.23.880}. 
Assume that the continuous position and momentum observables of subsystem $a(b)$ are $x^a (x^b)$ and $p^a (p^b)$, respectively. The states, in terms of continuous variables of momentum and position, satisfy the constraint 
\begin{align}
 h(p^b\vert p^a)+h(x^b\vert x^a)\geq \text{log}(\pi e),
 \end{align} 
 based on the entropic uncertainty relation (EUR) for the local hidden state model \cite{Bialynicki-Birula1975}. The above inequality is determined for discrete observables $R^a, S^a(R^b, S^b)$, taking into account the positivity of the relative Shannon entropy, such as \cite{PhysRevLett.60.1103} 
 \begin{align}
H(R^b\vert R^a)+H(S^b\vert S^a)\geq \text{log}(\Omega^b),
\end{align} 
with $H(R^b\vert R^a)\geq\sum_\lambda P(R^b\vert\lambda) H(R^b\vert\lambda)$, where $H(R^b\vert\lambda)$ represents the discrete Shannon entropy of probability distribution $P(R^b\vert\lambda)$  with hidden variable $\lambda$. Further, the eigenbasis of the corresponding marginal states are $\vert R_i \rangle$, $\vert S_j \rangle$ and $\text{log}(\Omega^b)=\text{min}_{ij}(1/\vert \langle R_i\vert S_j \rangle\vert)$. 
 The steering inequality for two-dimensional systems becomes  \cite{schneeloch}
 \begin{align}
    H(\sigma_x^b\vert\sigma_x^a)+ H(\sigma_y^b\vert\sigma_y^a)+ H(\sigma_z^b\vert\sigma_z^a)\geq 2,
\end{align}
when the Pauli $X$, $Y$, and $Z$ measurements basis is used. \textcolor{black}{Here, $H(\sigma_i^b\vert\sigma_i^a)$ for $i=x,y,z$ is the conditional Shannon entropy of Bob's outcomes given Alice's outcomes, reflecting on the EUR adapted to the steering scenario by replacing marginal entropies with conditional ones.} Further, the quantum steering is shown to be useful when the following inequality is violated for a general bipartite $X$-state $\rho$ \cite{Sun2017} as

\resizebox{.95\linewidth}{!}{
  \begin{minipage}{\linewidth}
\begin{align}
 QS = \sum_{j=1,2}\left[(1+c_j)\text{log}_2(1+c_j)+(1-c_j)\text{log}_2(1-c_j)\right]
  -  (1+r)\text{log}_2(1+r)+(\bar{r})\text{log}_2(\bar{r}) \nonumber\\
  +\frac{1}{2}(1+c_3+r+s)\text{log}_2(1+c_3+r+s) 
  +\frac{1}{2}(1+c_3-r-s)\text{log}_2(1+c_3-r-s) \nonumber \\
  +\frac{1}{2}(1-c_3-r+s)\text{log}_2(1-c_3-r+s) 
  +\frac{1}{2}(1-c_3+r-s)\text{log}_2(1-c_3+r-s)\leq 2, 
\end{align} 
\end{minipage}}
where, 
\begin{align} 
  c_1=2(\rho_{23}+\rho_{14}), ~~ c_2=2(\rho_{23}-\rho_{14}), ~~
  c_3=\rho_{11}+\rho_{44}-\rho_{22}-\rho_{33}, \\ \nonumber
  r=\rho_{11}+\rho_{22}-\rho_{44}-\rho_{33}, ~~
  s=\rho_{11}-\rho_{44}-\rho_{22}+\rho_{33}.
\end{align}
\textcolor{black}{This inequality was obtained by adapting the EUR for complete sets of mutually unbiased observables in two dimensions into the steering inequality by substituting conditional entropies for marginal ones, as detailed in Ref. \cite{Sun2017}.}

\section{Results and discussion} \label{Results}

The density matrix of a MEMS is given as
 \begin{align}
     \rho_{M} =
     \begin{pmatrix}
     g(\gamma) & 0 & 0 &\frac{\gamma}{2} \\
	0 & 1-2g(\gamma) & 0 &0\\
	0 & 0 & 0 & 0\\
	\frac{\gamma}{2} & 0 & 0 &g(\gamma)\\   
     \end{pmatrix},~\text{where}~~ 
     g (\gamma) =\begin{cases} \frac{1}{3}, &  0\le\gamma<\frac{2}{3}, \\
    \frac{\gamma}{2}, &  \frac{2}{3} \le\gamma\le 1.
    \end{cases} 
     \label{eq-MEMS}
     \end{align}
     The MEMS (\ref{eq-MEMS}) has the maximum amount of entanglement quantified by the entanglement of formation \cite{Conc_Wooters1998} for a given degree of impurity (or vice-versa) as measured by linearised entropy $S_L=4/3(1-\Tr[\rho^2])$ \cite{MEMS1_Munro2001}. For $\gamma=1$, Eq.~(\ref{eq-MEMS}) takes the form of a Bell-state. 
     
     The Werner state is given as
     \begin{align} 
     \rho_{W} = \frac{1}{4}(1-r_b)I_{4\times4}+r_b \rho_B, \label{Eq_WernerState}
     \end{align} 
     with Bell state $\rho_B=\left|\varphi\right>\left<\varphi\right|$ from $\left|\varphi\right> = \frac{1}{\sqrt{2}}(\left|00 \right>+\left|11 \right>)$ and $r_b$ determines the purity of the state. The $\rho_W$ is a pure state when $r_b=1$, whereas it is a maximally mixed state when $r_b=0$. \textcolor{black}{For different initial states $\rho$, the evolved density matrix $\rho_\text{WMR}[\rho]$ obtained from Eq. (11), and the closed-form expressions for Concurrence which was maximized to obtain optimal QMR strength $r$ are given in Appendix A, B, and C (for $\rho_B$, $\rho_W$, and $\rho_M$, respectively).}
    
    The three sets of plots provided in Figs.~(\ref{fig_BellState}), (\ref{fig_WernerState}), and (\ref{fig_MEMS}) detail the behavior of the various quantum correlations --- Dense Coding Capacity ($\chi$), Fidelity of teleportation ($\mathcal{F}$), Concurrence ($C$), EPR Steering or quantum steering (QS), Trace Distance Discord (TDD), and Jensen-Shannon divergence (JSD) --- under the influence of a CAD channel with the memory parameter (\(\eta\)) and the effect of one and two-qubit WMR on the Bell-state, Werner-state, and MEMS, respectively. 
    We consider decoherence strength as $p=0.5$ for all cases when QMR operation is applied by varying the WM strength $q$ with the corresponding optimized QMR strength $r$. 
    Each figure is structured as follows: (a) Evolution of the entangled state under pure AD channel (no memory), (b) WMR applied to one qubit under AD channel, (c) WMR applied to both qubits under AD channel, (d) Evolution under CAD channel (memory present, $\eta = 1$), (e) WMR applied to one qubit under CAD channel, (f) WMR applied to both qubits under CAD channel. \textcolor{black}{Since these correlations have different ranges, we plot the normalized correlations, N[correlation], for better comparison of the measures. With this normalization, values above the bold black line at zero separate the quantum bound for QS, $\chi$, and $\mathcal{F}$, which only exhibit useful behaviour above this threshold.} For instance, N[QS] is plotted, which is $\equiv$ (QS $-$ classical limit of QS)/(Maximum value of QS $-$ classical limit of QS). Maximum values and classical limits of $\chi$, $\mathcal{F}$, $C$, QS, TDD, JSD are 2, 1, 1, 6, 1, 0.56 and 1, 2/3, 0, 2, 0, 0, respectively.
     
\subsection{Bell-State}
\label{Bell-State}
     
\begin{figure}[htbp!]
    \centering
    \begin{minipage}[b]{0.32\textwidth}
        \centering
        \includegraphics[width=\textwidth]{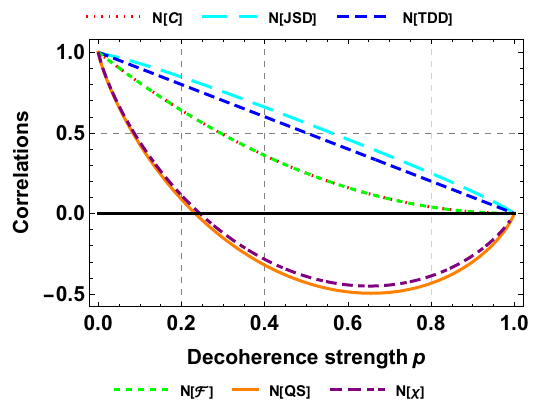} 
        \caption*{\small{(a) Correlations without WMR, $\eta=0$.}}
    \end{minipage}
    \begin{minipage}[b]{0.32\textwidth}
        \centering
        \includegraphics[width=\textwidth]{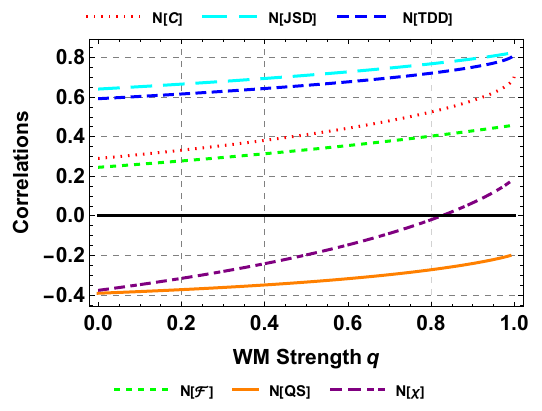} 
        \caption*{\small{(b) Correlations with WMR on one qubit, $\eta=0$, $p=0.5$.}}
    \end{minipage}
    \begin{minipage}[b]{0.33\textwidth}
        \centering
        \includegraphics[width=\textwidth]{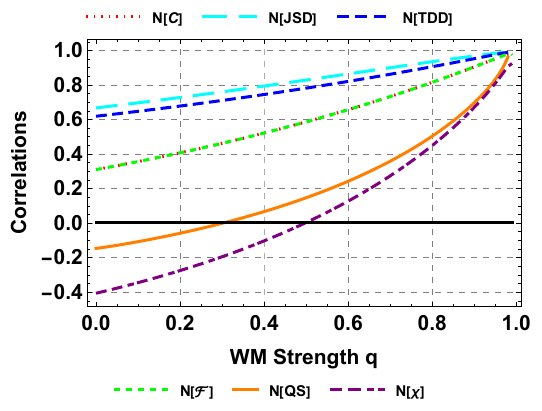} 
        \caption*{\small{(c) Correlations with WMR on two qubits, $\eta=0$, $p=0.5$.}}
    \end{minipage}
    
    \begin{minipage}[b]{0.32\textwidth}
        \centering
        \includegraphics[width=\textwidth]{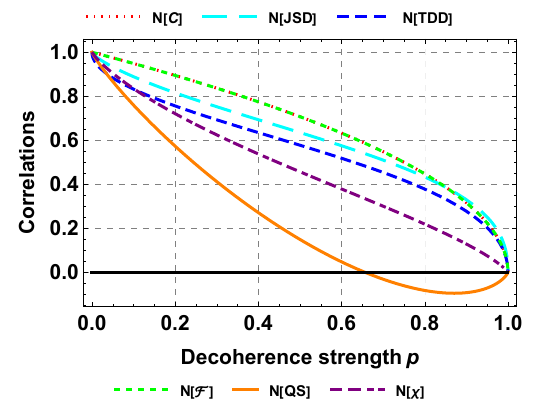} 
        \caption*{\small{(d) Correlations without WMR, $\eta=1$.}}
    \end{minipage}
    \begin{minipage}[b]{0.32\textwidth}
        \centering
        \includegraphics[width=\textwidth]{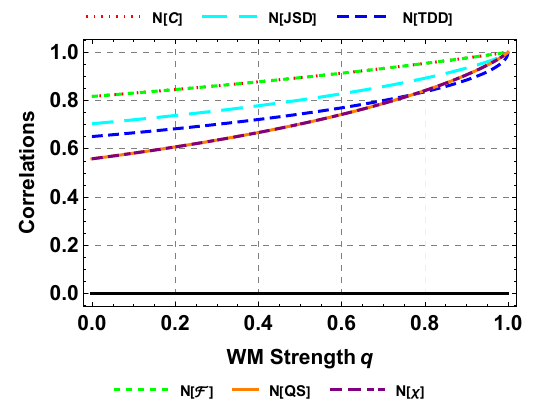} 
        \caption*{\small{(e) Correlations with WMR on one qubit, $\eta=1$, $p=0.5$.}}
    \end{minipage}
    \begin{minipage}[b]{0.33\textwidth}
        \centering
        \includegraphics[width=\textwidth]{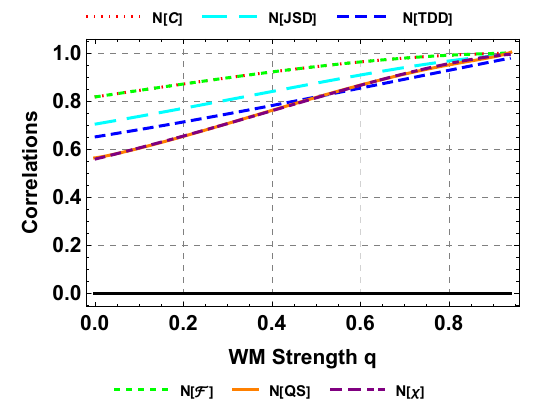} 
        \caption*{\small{(f) Correlations with WMR on two qubits, $\eta=1$, $p=0.5$.}}
    \end{minipage}
    \caption{\textit{Quantum correlations for the Bell state (Werner state with $r_b = 1$, and MEMS with $\gamma = 1$). (a) and (d) show the effect of the CAD channel's decoherence parameter $p$ with correlation parameters $\eta=0$ and $\eta=1$, respectively. (b) and (e) show the effect of the one-qubit WMR protocol, whereas (c) and (f) show the effect of the two-qubit WMR protocol at $p=0.5$ in the absence ($\eta=0$) and presence ($\eta=1$) of memory, respectively, with $q$ being the weak measurement strength.}}
    \label{fig_BellState}
\end{figure}

The initial Bell state in Fig.~(\ref{fig_BellState}) is maximally entangled and, as such, exhibits stronger quantum correlations compared to the Werner states and MEMS. \textcolor{black}{The measures N[QS], N[$C$], N[TDD], and N[JSD] represent different types of quantum correlations. Yet, they exhibit a hierarchical behaviour as discussed in Section ~(\ref{Introduction}). 
Stronger measures like N[QS] (and similarly N[$\chi$]) are most sensitive to decoherence, while weaker measures of N[TDD] and N[JSD] exhibit a gradual degradation showing lower sensitivity to decoherence. The measures N[$C$] (and similarly N[$\mathcal{F}$]) display an intermediate behaviour. Their decay and reversal behaviour generally indicates that while they capture different aspects of quantumness, their qualitative nature under decoherence is consistent with the hierarchy.  Notably, the behaviour of these correlations in the classical regime, i.e., below zero, is physically inconsequential.} 

 Without memory, N[JSD] and N[TDD] persist in decay (Fig.~(\ref{fig_BellState}a)), and show fast recovery under reversal (Figs.~(\ref{fig_BellState}b), (c)) indicating that some quantum correlations survive even as entanglement (N[$C]$) and teleportation N[$\mathcal{F}]$ show diminished behaviour. 
 In contrast, N[$\mathcal{F}$] and N[$C$] persist in decay and reversal primarily with the memory case (Figs.~(\ref{fig_BellState}d), (e), (f)). 
  \textcolor{black}{Also, with memory, N[TDD] and N[JSD] display intermediate behavior towards decoherence, while N[TDD] shows a non-monotonic crossover behaviour with N[QS] and N[$\chi$]}. 
 Without memory, N[QS] and N[$\chi$] are the most vulnerable to noise as they lose the quantum advantage close to $p \approx 0.25$, indicating the loss of steering and quantum-dense coding at moderate levels of decoherence. 
  In the presence of memory, while all correlations show enhanced robustness to noise as compared to the ADC, some measures decay more steeply than others (Figs.~(\ref{fig_BellState}a), (d)). This steepness reflects varying sensitivity to decoherence. For example, N[QS] continues to undergo sudden death, but at a delayed $p\
  \approx 0.66$, and N[$\chi$] exhibits an asymptotic decay compared to its sudden death behaviour in the ADC. $N[\chi]$ and N[QS] drop below zero (the classical limit) and then asymptotically approach the value zero as $p\rightarrow 1$ as the state becomes pure again. 
 At the end of the decoherence process, the two qubits land in a pure separable state $|00\rangle$. This behaviour can be correlated with the purity of the state defined as Tr($\rho^2$), previously observed in Fig.~(5.16) of Ref. \cite{Thesis_Singh2022}. 
  
  With WMR on one qubit and no memory (Fig.~\ref{fig_BellState}b), all correlations except N[QS] see some recovery as $q$ increases. 
  With WMR on both qubits and no memory (Fig.~\ref{fig_BellState}c), the correlations recover strongly as compared to the one-qubit WMR case, approaching their maximum values as $q\rightarrow 1$. Notably, N[QS] and N[$\chi$] recover prominently and rise above zero for a larger range of $q$, ensuring that the state retains its nonlocal steering and dense coding capabilities. 
  Memory combined with WMR on one qubit (Fig.~\ref{fig_BellState}e) gives better recovery compared to (Fig.~\ref{fig_BellState}c), with all correlations returning faster to their maximum values as $q\rightarrow 1$. 
  WMR on both qubits with memory (Fig.~\ref{fig_BellState}f) yields the best correlation recovery. Correlations approach their maximum values asymptotically, ensuring all quantum correlations remain robust against decoherence. 
  While both N[$\mathcal{F}$] and N[$\chi$] are quantum communication applications utilizing entanglement, N[$\mathcal{F}$] shows better resilience against the noise as compared to N[$\chi$].

\subsection{Werner-State}

\begin{figure}[htbp!]
    \centering
    \begin{minipage}[b]{0.32\textwidth}
        \centering
        \includegraphics[width=\textwidth]{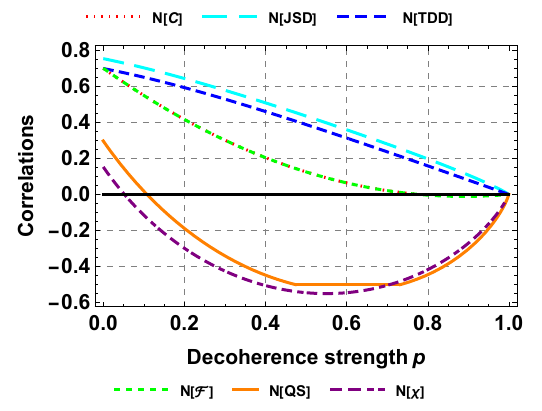} 
        \caption*{\small{(a) Correlations without WMR,  $\eta=0$.}}
    \end{minipage}
    \begin{minipage}[b]{0.32\textwidth}
        \centering
        \includegraphics[width=\textwidth]{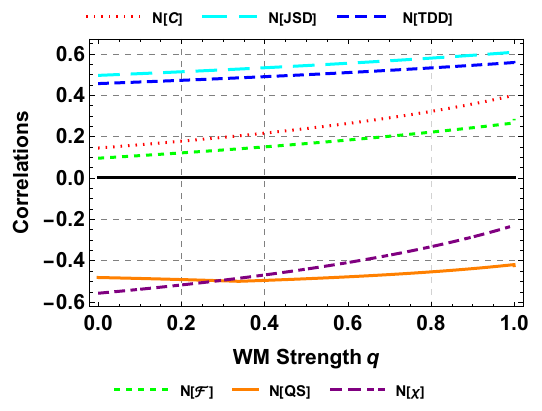} 
        \caption*{\small{(b) Correlations with WMR on one qubit, $\eta=0$, $p=0.5$.}}
    \end{minipage}
    \begin{minipage}[b]{0.33\textwidth}
        \centering
        \includegraphics[width=\textwidth]{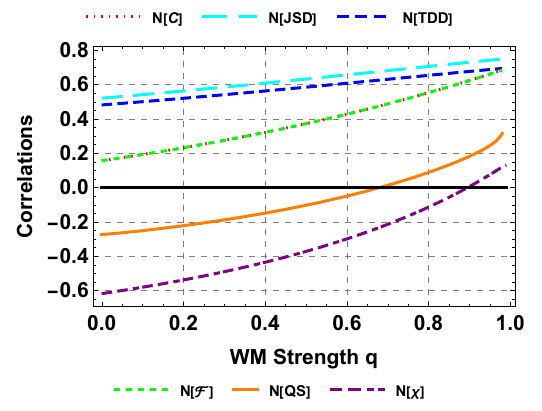} 
        \caption*{\small{(c) Correlations with WMR on two qubits, $\eta=0$, $p=0.5$.}}
    \end{minipage}
    
    \begin{minipage}[b]{0.32\textwidth}
        \centering
        \includegraphics[width=\textwidth]{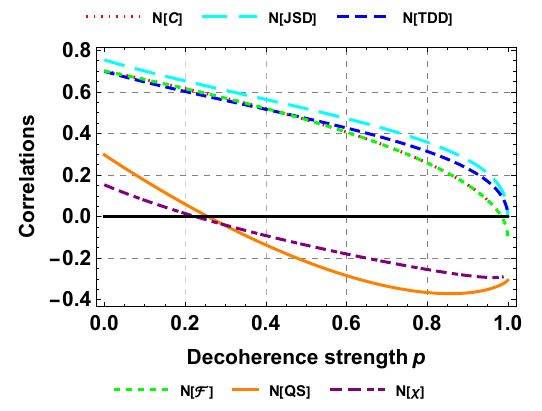} 
        \caption*{\small{(d) Correlations without WMR, $\eta=1$.}}
    \end{minipage}
    \begin{minipage}[b]{0.32\textwidth}
        \centering
        \includegraphics[width=\textwidth]{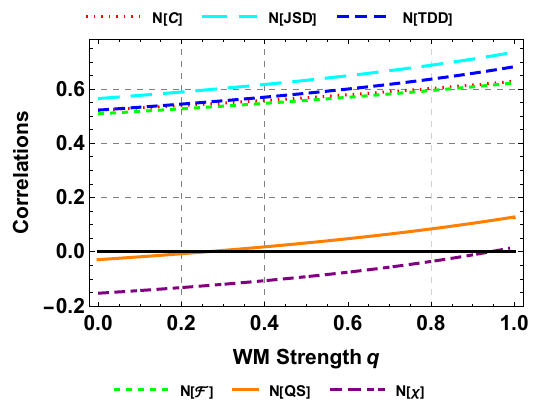} 
        \caption*{\small{(e) Correlations with WMR on one qubit, $\eta=1$, $p=0.5$.}}
    \end{minipage}
    \begin{minipage}[b]{0.33\textwidth}
        \centering
        \includegraphics[width=\textwidth]{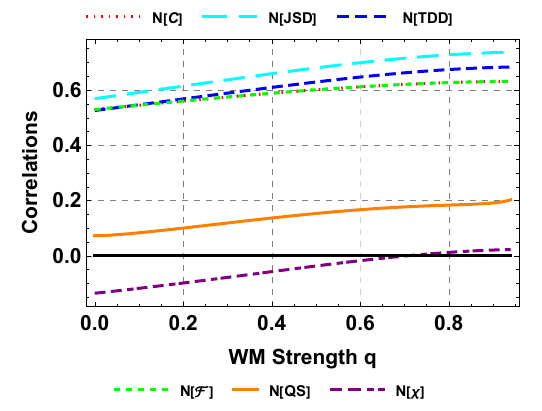} 
        \caption*{\small{(f) Correlations with WMR on two qubits, $\eta=1$, $p=0.5$.}}
    \end{minipage}
    \caption{\textit{Quantum correlations for the Werner state with $r_b=0.8$. (a) and (d) show the effect of the CAD channel's decoherence parameter $p$ with correlation parameters $\eta=0$ and $\eta=1$, respectively. (b) and (e) show the effect of the one-qubit WMR protocol, whereas (c) and (f) show the effect of the two-qubit WMR protocol at $p=0.5$ in the absence ($\eta=0$) and presence ($\eta=1$) of memory, respectively, with $q$ being the weak measurement strength}.}
    \label{fig_WernerState}
\end{figure}

The Werner state (Eq.~\ref{Eq_WernerState}) with $r_b=0.8$ is partially entangled, and therefore the state loses quantum advantage faster than a Bell state as shown in Fig.~(\ref{fig_WernerState}). Without WMR and no memory (Fig.~(\ref{fig_WernerState}a)),  correlations decay quickly as $p$ increases compared to the Bell-state $r_b=1$ case. The N[$\chi$] and N[QS] rapidly fall below zero (classical limit). N[JSD] and N[TDD] decrease but remain nonzero for $p\in [0,1)$, suggesting some persistence of nonclassical correlations even as other measures of dense coding, steering, entanglement, and teleportation are lost in a sudden death manner. Without WMR, the presence of memory in the CAD channel slows down the decay of all correlations (Fig.~(\ref{fig_WernerState}d)). The N[$\chi$] and N[QS], which were most vulnerable in the ADC, remain above zero for longer as $p$ increases, indicating a slightly better resilience of dense coding and steering due to the memory effect. \textcolor{black}{The hierarchy of the correlations shows similar behaviour as discussed for the Bell state.}

Applying WMR (Figs.~ (\ref{fig_WernerState}b), (c), (e), (f)) helps the correlations partially recover with coherence (N[JSD]) and discord (N[TDD]) recovering more quickly than entanglement (N[$C$]) and teleportation (N[$\mathcal{F}$]) as $q$ increases.
However, when WMR is applied to just one qubit, N[$\chi$] and N[QS] show no recovery, indicating that dense coding and steering are challenging to restore without memory (Figs.~(\ref{fig_WernerState}b)) as compared to with memory case (Fig.~(\ref{fig_WernerState}e)). 
Combining memory with WMR on one qubit further enhances all other correlations. N[QS] and N[$\chi$] improve more, reaching above zero for larger values of $q$. 
Applying WMR to both qubits without memory shows a more robust recovery across all measures (Fig.~ (\ref{fig_WernerState}c)). Also, N[$\chi$] and N[QS] show a notable rise above the zero thresholds, becoming meaningful at higher values of $q$, suggesting greater robustness to decoherence. WMR on both qubits with memory gives the best results for the Werner state, with correlations recovering dramatically as $q$ increases and remaining robust even for more considerable decoherence strengths (Fig.~ (\ref{fig_WernerState}f)). N[QS] stays well above zero, ensuring that steering remains possible across a broad range of $q$, while N[$\chi$] recovers slightly only for larger values of $q>0.7$. 

\newpage 
\subsection{Maximally Entangled Mixed State (MEMS)}

\begin{figure}[htbp!]
    \centering
    \begin{minipage}[b]{0.31\textwidth}
        \centering
        \includegraphics[width=\textwidth]{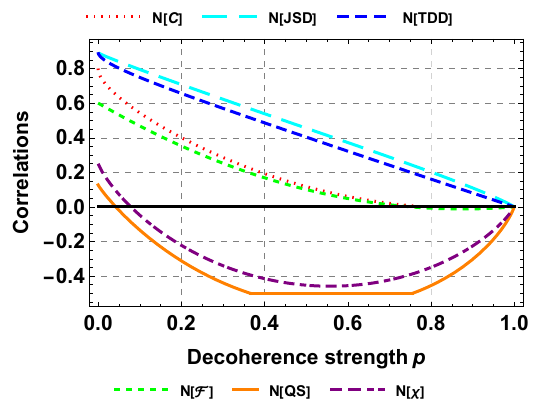} 
        \caption*{\small{(a) Correlations without WMR, $\eta=0$.}}
    \end{minipage}
    \begin{minipage}[b]{0.32\textwidth}
        \centering
        \includegraphics[width=\textwidth]{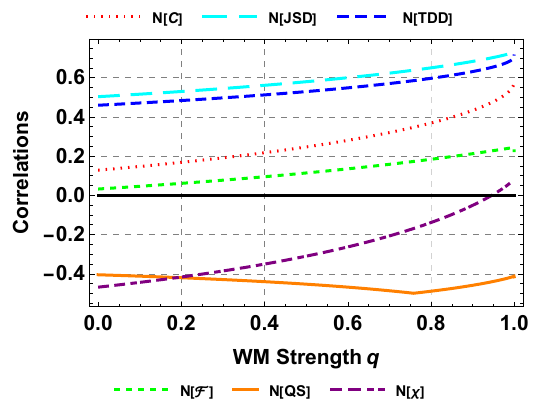} 
        \caption*{\small{(b) Correlations with WMR on one qubit, $\eta=0$, $p=0.5$.}}
    \end{minipage}
    \begin{minipage}[b]{0.33\textwidth}
        \centering
        \includegraphics[width=\textwidth]{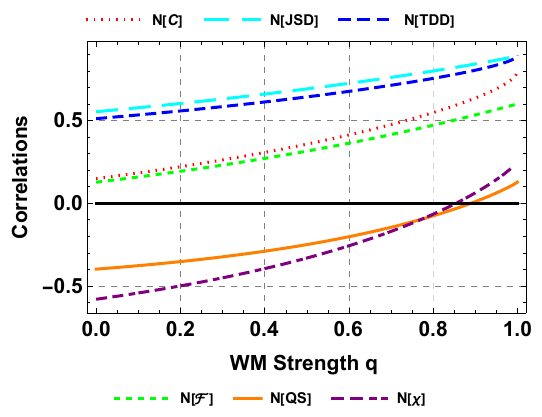} 
        \caption*{\small{(c) Correlations with WMR on two qubits, $\eta=0$, $p=0.5$.}}
    \end{minipage}
    
    \begin{minipage}[b]{0.32\textwidth}
        \centering
        \includegraphics[width=\textwidth]{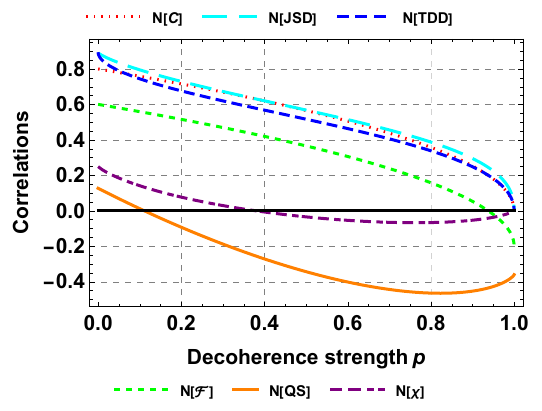} 
        \caption*{\small{(d) Correlations without WMR, $\eta=1$.}}
    \end{minipage}
    \begin{minipage}[b]{0.32\textwidth}
        \centering
        \includegraphics[width=\textwidth]{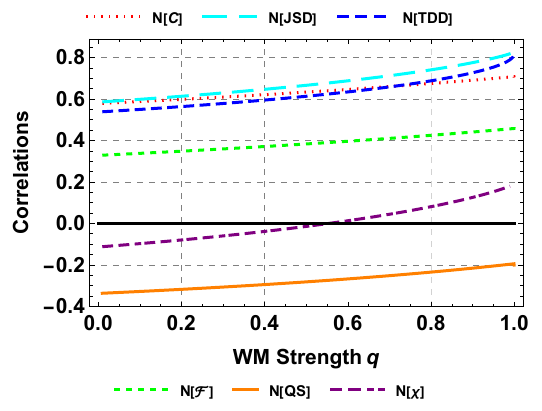} 
        \caption*{\small{(e) Correlations with WMR on one qubit, $\eta=1$, $p=0.5$.}}
    \end{minipage}
    \begin{minipage}[b]{0.33\textwidth}
        \centering
        \includegraphics[width=\textwidth]{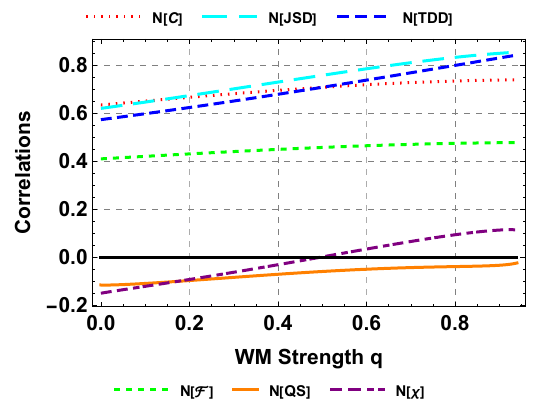} 
        \caption*{\small{(f) Correlations with WMR on two qubits, $\eta=1$, $p=0.5$.}}
    \end{minipage}
    \caption{\textit{Correlations for the MEMS with $\gamma=0.8$. (a) and (d) show the effect of the CAD channel's decoherence parameter $p$ with correlation parameters $\eta=0$ and $\eta=1$, respectively. (b) and (e) show the effect of the one-qubit WMR protocol, whereas (c) and (f) show the effect of the two-qubit WMR protocol at $p=0.5$ in the absence ($\eta=0$) and presence ($\eta=1$) of memory, respectively, with $q$ being the weak measurement strength.}}
    \label{fig_MEMS}
\end{figure}

The MEMS (Eq.~\ref{eq-MEMS}) with $\gamma=0.8$ starts with an initial entanglement higher than the Werner state ($r_b=0.8$) but less than the Bell state. Without WMR in AD and CAD channels, all correlations decay rapidly as $p$ increases. As shown in Figs.~(\ref{fig_MEMS}a) and (\ref{fig_MEMS}d), N[QS] quickly falls below zero (followed by N[$\chi$]), showing that quantum steering is most sensitive to the noise in MEMS. N[TDD] and N[JSD] persist for longer and show asymptotic decay compared to N[$C$] and N[$\mathcal{F}$], which show sudden death (similar to the Werner state). Memory helps slow the decay, but N[QS] and N[$\chi$] still fall below zero at smaller values of $p$ compared to the Bell state. 

With WMR on one qubit (Figs.~(\ref{fig_MEMS}b), (e)), some recovery of all correlations (except for N[QS]) is seen as $q$ increases.
WMR on both qubits without memory (Fig.~(\ref{fig_MEMS}c)) helps recover all the correlations, while with memory (Fig.~(\ref{fig_MEMS}f)), N[QS] does not recover at all.  This observation contrasts with the Werner state, where all correlations show some recovery for WMR on both qubits. WMR on two qubits without memory (Fig.~(\ref{fig_MEMS}c)) yields the best recovery for all correlations in MEMS.
Also, in Figs.~(\ref{fig_MEMS}c), N[$C$] and N[$\mathcal{F}$] recover to the initial values, whereas in all other WMR cases (with and without memory), N[$C$] recovery is comparatively better than N[$\mathcal{F}$].
We assume that the composition of the MEMS makes the behaviour of N[$C$] and N[$\mathcal{F}$] differ compared to their nearly same behaviour in plots of Figs.~(\ref{fig_BellState}) and (\ref{fig_WernerState}). 

\subsection{\textcolor{black}{Non-maximally entangled state}}

\begin{figure}[htbp!]
    \centering
    \begin{minipage}[b]{0.49\textwidth}
        \centering
        \includegraphics[width=\textwidth]{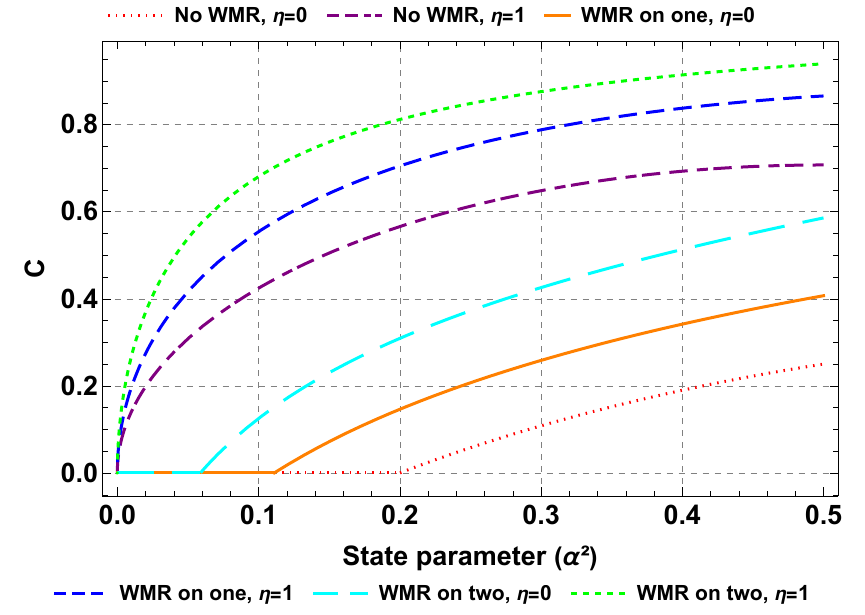} 
        \caption*{\small{(a) \textcolor{black}{Concurrence ($C$) vs $|\alpha|^2$.}}}
    \end{minipage}
    \begin{minipage}[b]{0.49\textwidth}
        \centering
        \includegraphics[width=\textwidth]{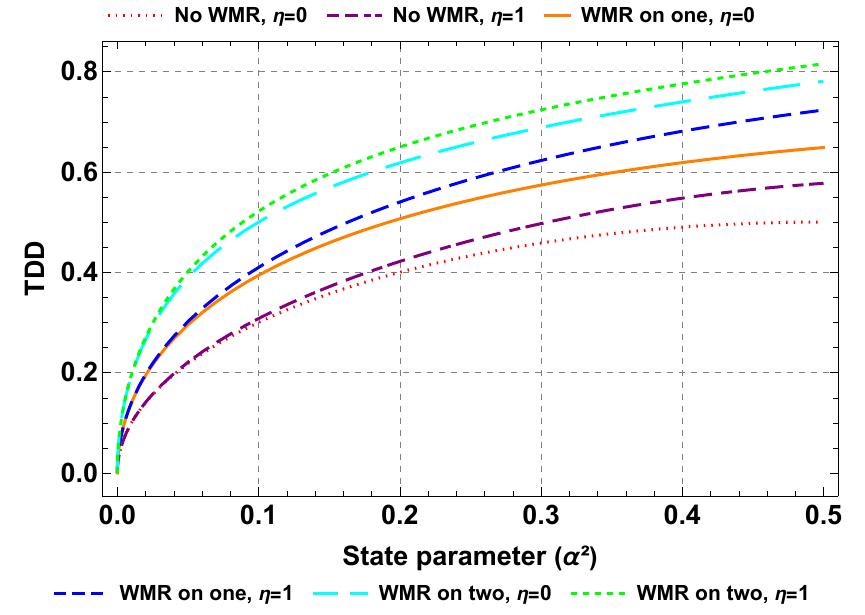} 
        \caption*{\small{(b) \textcolor{black}{Trace Distance Discord (TDD) vs $|\alpha|^2$.}}}
    \end{minipage}
    \caption{\textcolor{black}{\textit{Concurrence ($C$) and TDD for the state $|\psi\rangle=\alpha \ket{00} + \beta \ket{11}$ for different $\alpha$. For no WMR and memory parameter $\eta=0,1$, the decoherence parameter is set to $p=0.5$. For WMR on one or two qubits with $\eta=0,1$, $p=0.5$, WM strength is set to $q=0.5$, and optimal QMR strength $r$ is found through numerical optimization.}}}
    \label{fig_NMES}
\end{figure}

\textcolor{black}{To offer a different perspective on the above results while highlighting the advantage of two-qubit WMR over one-qubit WMR and no WMR (with or without memory), we compute the concurrence and TDD for non-maximally entangled states of the form $|\psi\rangle = \alpha \ket{00} + \beta \ket{11}$ by varying the state parameter $\alpha$, where $|\alpha|^2 + |\beta|^2 = 1$.
We study these correlations for non-maximally entangled states $\rho=|\psi\rangle\langle\psi|$ under CAD channel ($\eta=0,1$, no WMR) evolution as given in Eq. (7) with decoherence strength $p=0.5$. Subsequently, the effect of the WMR operation (WM followed by CAD and then QMR) on one and two qubits is studied for $p=0.5$, with moderate WM strength $q=0.5$, and corresponding optimal QMR strength $r$ as given in Eq. (11) (for $\eta=0,1$). These results are shown in Fig.~(\ref{fig_NMES}).
Observing the $\eta=0$ and $\eta=1$ case separately in Fig. (\ref{fig_NMES}), correlations $C$ and TDD clearly show the advantage of WMR on two-qubits compared to just one.
However, when comparing the effect of the memory parameter $\eta$ with WMR operation, the results differ for  $C$ and TDD. Specifically, in the context of $C$, the WMR on one qubit with $\eta=1$ resp., No WMR, $\eta=1$) is more advantageous than WMR on two qubits with $\eta=0$ resp.,  WMR on one qubit, $\eta=0$). In contrast, TDD shows just the opposite behaviour as shown in Fig.~(\ref{fig_NMES}) plots (a) and (b). This, and other similar subtle differences, emphasize the importance of the different correlation measures used in this study, despite their broadly similar behaviour.}

\section{Analysis with neural networks}  
\label{ANN}

A neural network is a computational model comprising layers of artificial neurons that process information, much like the human brain, by processing impulses. Each neuron takes in a set of inputs as information and uses activation functions, weights, and biases to decide its output. For the artificial neural network to provide predictions or classifications, these weights are adjusted by a learning algorithm that assesses the input and output data. The network learns the relationship between these inputs and outputs through back-propagation, which iteratively reduces the error between expected and actual outputs. We use Fitnet, a MATLAB fitting neural network function \cite{fitnet}, to learn and predict direct TDD values using other unnormalized correlations directly as input. In this network, a particular weight connects each input (namely, JSD, Concurrence, Fidelity of teleportation, EPR steering, and capacity of dense coding) to the neurons of the artificial neural network. We determine these weights, which are crucial in determining how much influence each input feature has on the output TDD predictions made by the network. We use this to study the relationship between the other correlations and quantum discord.

\begin{figure}[htbp!]
    \centering
    \includegraphics[width=1\linewidth]{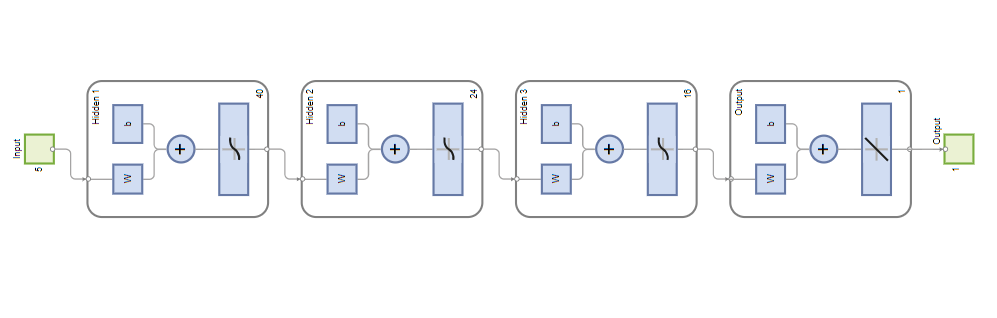}
    \caption{\textit{MATLAB neural network model consisting of 80 neurons with three hidden layers.}}
    \label{fig_MatlabANN}
\end{figure}

As shown in Fig.~(\ref{fig_MatlabANN}), we chose a neural network with 80 neurons divided into three hidden layers of $40,24,16$ each. We found that deep networks (with more than two layers) can allow more complex relationships in data than shallow networks (which only have one or two hidden layers). Larger models can be achieved by packing more neurons into fewer layers, although overfitting is a risk, particularly when the data is scarce. The network may eventually abstract information and learn more intricate patterns by adding layers. The training function used is the Levenberg-Marquardt algorithm. It is designed specifically for small to medium-sized problems and offers rapid convergence.

\textcolor{black}{The three neural network layers were assigned different activation functions to optimize the learning process. The first layer uses a log-sigmoid activation function, which is often preferred in the initial layer as it helps simulate probabilities and introduces non-linearity early in the network. This function maps inputs into a constrained range, improving stability and gradient updates in deeper layers. In the second hidden layer, we employ the tangent-sigmoid activation function, which enhances non-linearity further and provides a steeper gradient for training. This layer is crucial for capturing complex interactions between quantum correlations. Finally, the third hidden layer employs a linear activation function, as linear regression is ideal for output layers in continuous-valued predictions. This choice ensures the model can directly map learned features to a continuous output space corresponding to the predicted TDD values.}

\textcolor{black}{The network's weight adjustments during training provide insight into the relative importance of each input correlation in predicting TDD. Since each input correlation (such as JSD, Concurrence, or EPR steering, different correlations in the hierarchy) is assigned a unique weight, analyzing these values allows us to quantify their contributions to the TDD prediction. This enables a deeper understanding of the interplay between quantum correlations and quantum discord.}

\textcolor{black}{The advantage of this network architecture is that while increasing the number of layers and neurons can improve the model's ability to learn complex relationships, excessive depth or neuron count can lead to overfitting, where the model memorizes noise instead of learning meaningful patterns. We control complexity while ensuring hierarchical feature extraction by carefully selecting three hidden layers with decreasing neuron counts (40, 24, and 16). The first layer captures broad patterns, the second layer refines these relationships, and the third layer ensures a smooth transition to output prediction.}

 The neural network prediction performance is assessed using Mean Squared Error (MSE). The MSE calculates the average deviations between the expected TDD and the actual TDD values. It is provided by \begin{align}
\text{MSE} = \frac{1}{n} \sum_{i=1}^{n} ( T_i - \hat{T_i} )^2,
\end{align} where $\hat{T_i}$ is the predicted value for $n$ predictions and $T_i$ is the actual value of TDD. Lower MSE values indicate better performance because MSE depends on the data's scale; however, there is no ideal MSE value. In general, a lower MSE value is preferred. We ran over 20 iterations of the neural network predictions and chose the instance with the lowest MSE value. We study the TDD prediction only for scenarios without WMR and WMR applied to two qubits. The network learns the weights that link inputs to the hidden layer differently in each run. These weights indicate each input feature's relative relevance. A weight's magnitude indicates how much the related input contributed to the network's prediction. The average relevance of each input is shown in a bar plot, with bars showing variation between all the neurons. This determines each input feature's overall impact on every neuron in the first hidden layer.

\begin{figure}[H]
    \centering
    \begin{minipage}[b]{0.24\textwidth}
        \centering
        \includegraphics[width=\textwidth]{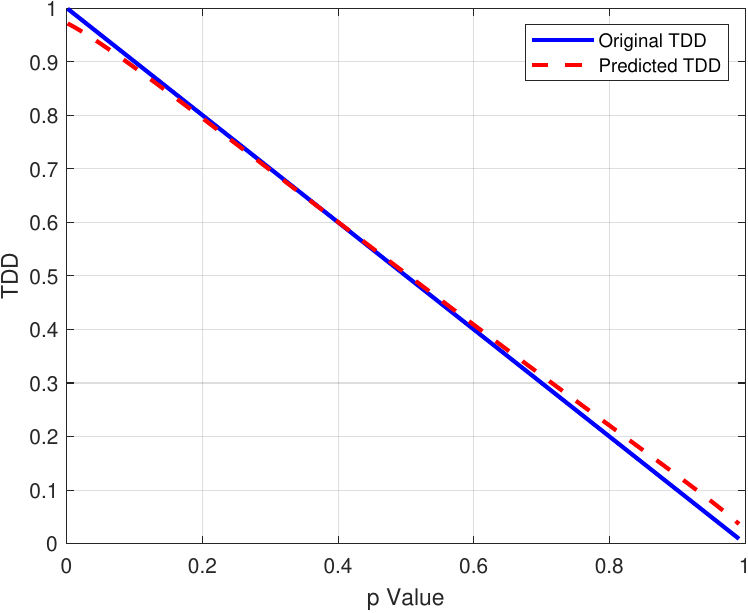} 
        \caption*{(a) \small{Without WMR; $\eta=0$, MSE $=2.3\times 10^{-4}$.}}
    \end{minipage}
    \begin{minipage}[b]{0.24\textwidth}
        \centering
        \includegraphics[width=\textwidth]{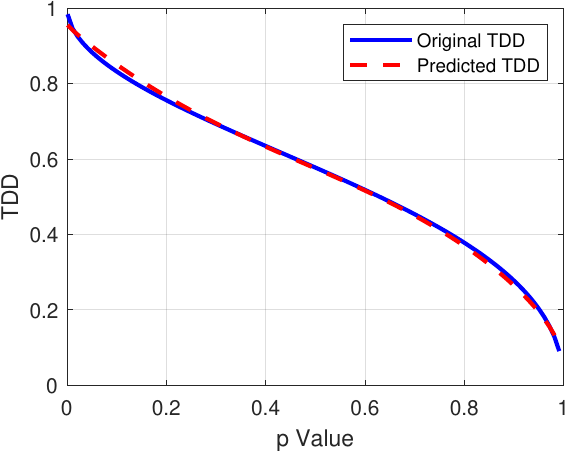} 
        \caption*{\small{(b) Without WMR; $\eta=1$, MSE $=8.5\times10^{-5}$.}}
    \end{minipage}
    \begin{minipage}[b]{0.24\textwidth}
        \centering
        \includegraphics[width=\textwidth]{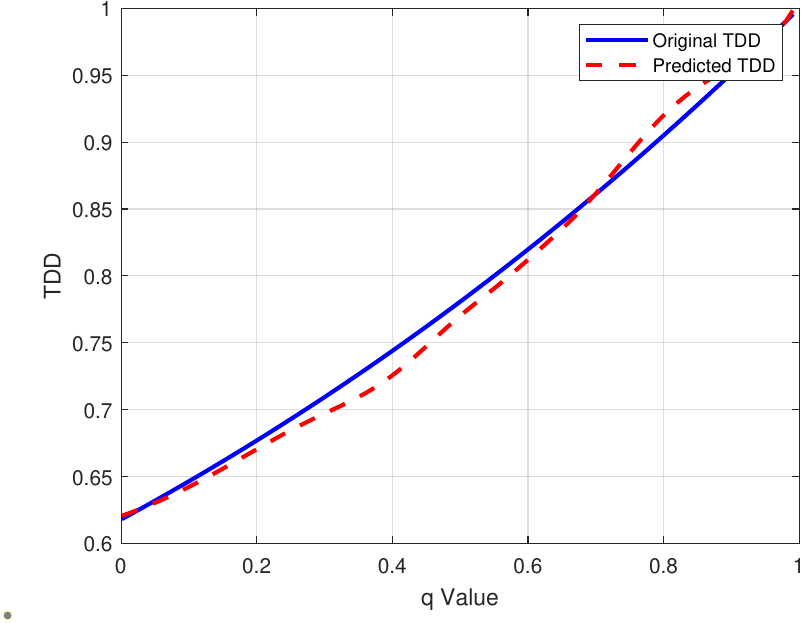} 
        \caption*{ \small{(c) WMR on both qubits; $\eta=0$, MSE $=9.8\times10^{-5}$.}}
    \end{minipage}
    \begin{minipage}[b]{0.24\textwidth}
        \centering
        \includegraphics[width=\textwidth]{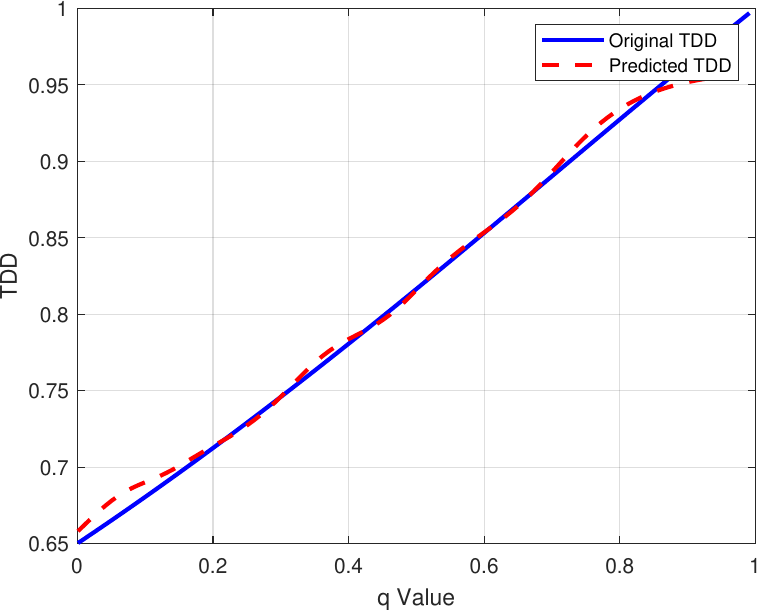} 
        \caption*{\small{(d) WMR on both qubits; $\eta=1$, MSE $=1\times10^{-4}$.}}
    \end{minipage}
     \caption{\textit{Neural network predictions for TDD from JSD, Concurrence, teleportation Fidelity, EPR steering, and capacity of dense coding for Bell state. The dashed line shows the prediction against the actual value of TDD in a continuous line.}}
     \label{fig_TDD-BellState}
\end{figure}

\begin{figure}[htb!]
    \centering
    \begin{minipage}[b]{0.24\textwidth}
        \centering
        \includegraphics[width=\textwidth]{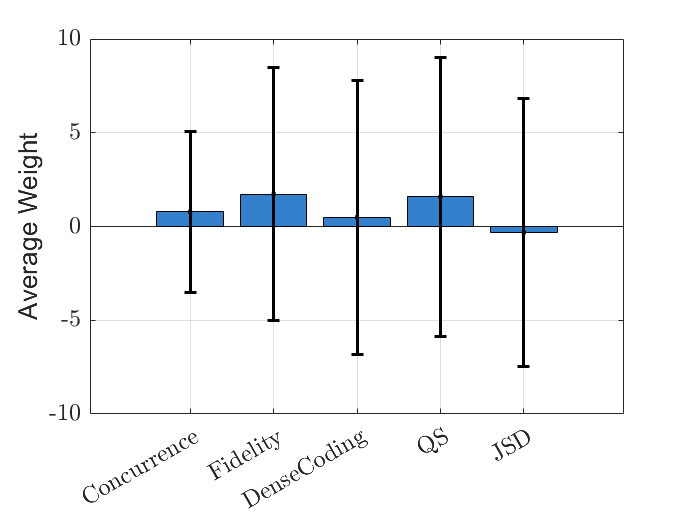} 
        \caption*{(a) Without WMR, \newline $\eta=0$.}
    \end{minipage}
    \begin{minipage}[b]{0.24\textwidth}
        \centering
        \includegraphics[width=\textwidth]{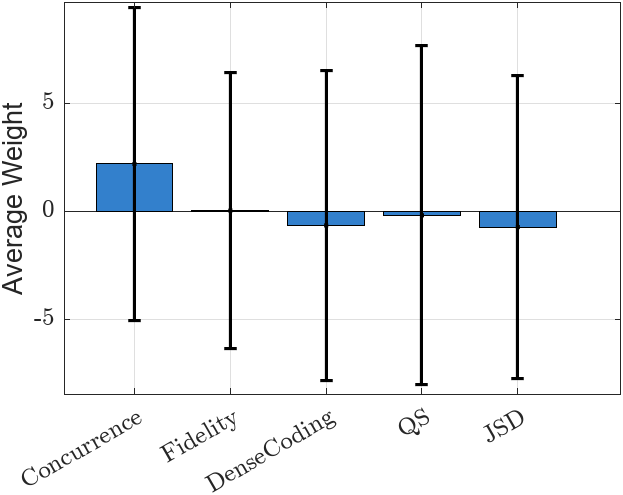} 
        \caption*{(b) Without WMR, \newline $\eta=1$.}
    \end{minipage}
    \begin{minipage}[b]{0.24\textwidth}
        \centering
        \includegraphics[width=\textwidth]{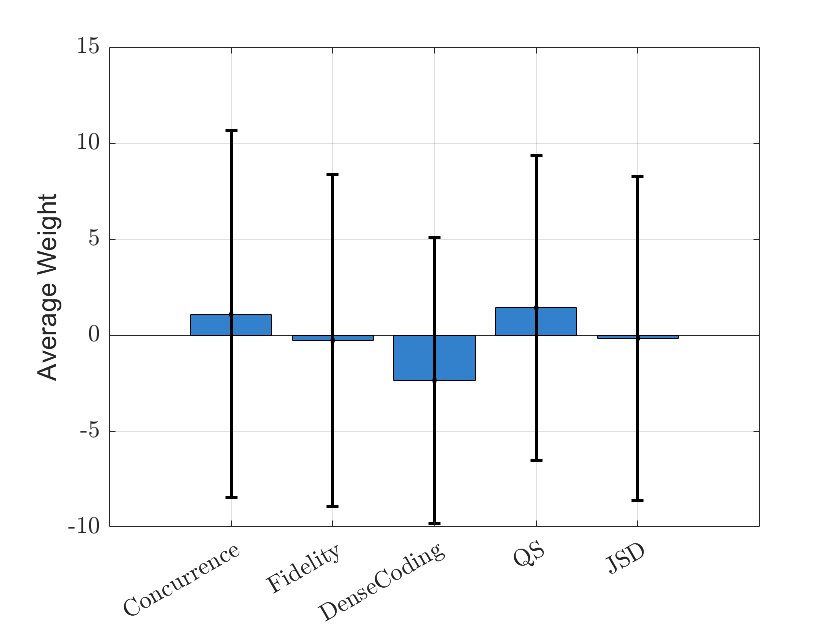} 
        \caption*{(c) WMR on both qubits, $\eta=0$.}
    \end{minipage}
    \begin{minipage}[b]{0.24\textwidth}
        \centering
        \includegraphics[width=\textwidth]{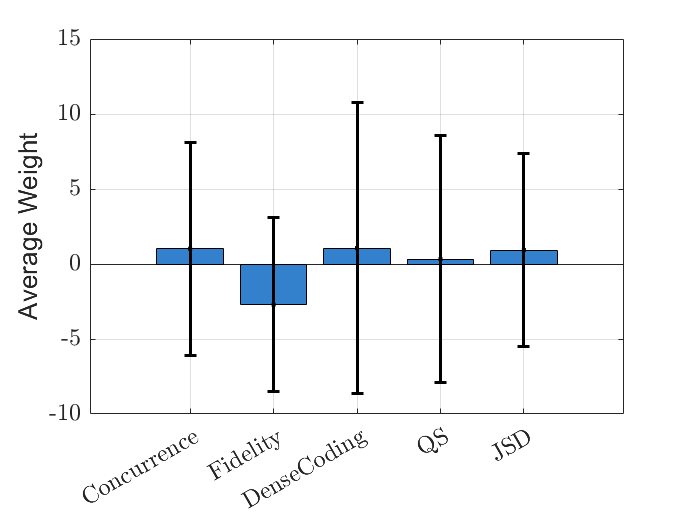} 
        \caption*{(d) WMR on both qubits, $\eta=1$.}
    \end{minipage}
     \caption{\textit{Average weights of neurons in the first layer with bars showing variations over different neurons in the prediction of TDD from JSD, Concurrence, teleportation Fidelity, EPR steering, and capacity of dense coding for Bell state as input.}}
     \label{fig_TDDWeights-BellState}
\end{figure}

For the perfectly entangled Bell state studied in Fig.~(\ref{fig_BellState}), we observe that Concurrence generally has a strong positive influence on the TDD predictions, as seen from the neural network weights visualized in Fig.~(\ref{fig_TDDWeights-BellState}). The first hidden layer, consisting of 40 neurons, displays a broad distribution of positive and negative weights. This indicates that each neuron extracts distinct patterns from the input features, contributing uniquely to the final output. As a result, the overall network learns a comprehensive representation of how each quantum correlation contributes to predicting the trace distance discord. \textcolor{black}{In the scenario where both qubits are subjected to the WMR process without memory (Fig.~\ref{fig_TDDWeights-BellState}c), the dense coding capacity $\chi$ shows the most significant negative influence on TDD prediction. This aligns with Fig.~(\ref{fig_BellState}c), where TDD and $\chi$ evolve occupying the extremes of the correlation spectrum. Conversely, in the case where both qubits undergo WMR with memory (Fig.~\ref{fig_TDDWeights-BellState}d), the Fidelity $\mathcal{F}$ exerts the most substantial negative influence on the prediction. This is reflected in Fig.~(\ref{fig_BellState}f), where $\mathcal{F}$ and TDD lie on the extremes of the correlation spectrum. As discussed in Sec.~(3.1), the robustness of $\mathcal{F}$ increases with memory. Interestingly, we also observe a shift from positive to negative influence in the weights associated with $\mathcal{F}$, quantum steering (QS), and Jensen-Shannon divergence (JSD), as seen when comparing Fig.~(\ref{fig_TDDWeights-BellState}) plots (a) and (b). This transition may be attributed to a change in the relative sensitivity to decoherence of correlations under the memory parameter.} Samples of the predicted TDD values are shown in Fig.~(\ref{fig_TDD-BellState}). The close agreement between predicted and actual values and low MSE confirms that the neural network effectively captures the underlying relationships between the input correlations and TDD.

\textcolor{black}{For the mixed entangled Werner state ($r_b=0.8$) studied in Fig.~(\ref{fig_WernerState}), we observe that primarily QS has a positive influence. Concurrence has mostly a negative influence from the weights visualized in Fig.~(\ref{fig_TDDWeights-Wstate}) (from the previous section, we can observe that entanglement behaviour matches less with TDD behaviour for the less entangled state in plots of Fig.~(\ref{fig_WernerState}). The Fidelity $\mathcal{F}$ also has a positive impact. In contrast, the dense coding capacity again has the most negative influence on the prediction of TDD. JSD positively influences TDD prediction in this case, without WMR, contrary to the previous Bell state case.} Samples of the TDD prediction are shown in Fig.~(\ref{fig_TDDWeights-BellState}). The predicted correlations of TDD in this less entangled state match less closely than in the perfectly entangled cases with the actual values.

\begin{figure}[htb!]
    \centering
    \begin{minipage}[b]{0.24\textwidth}
        \centering
        \includegraphics[width=\textwidth]{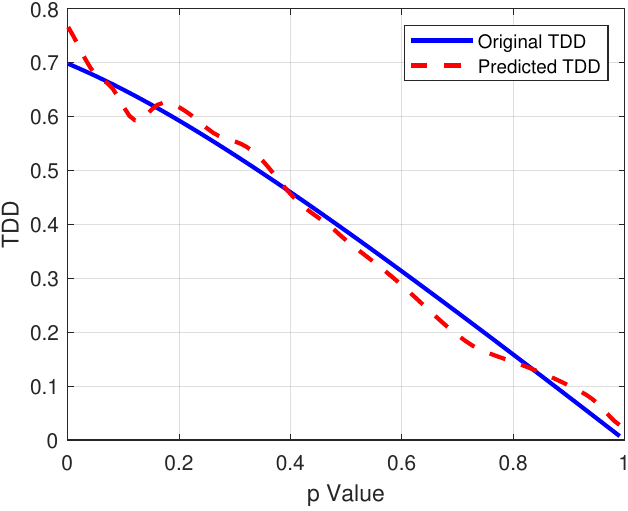} 
        \caption*{(a) Without WMR, \newline $\eta=0$.}
    \end{minipage}
    \begin{minipage}[b]{0.24\textwidth}
        \centering
        \includegraphics[width=\textwidth]{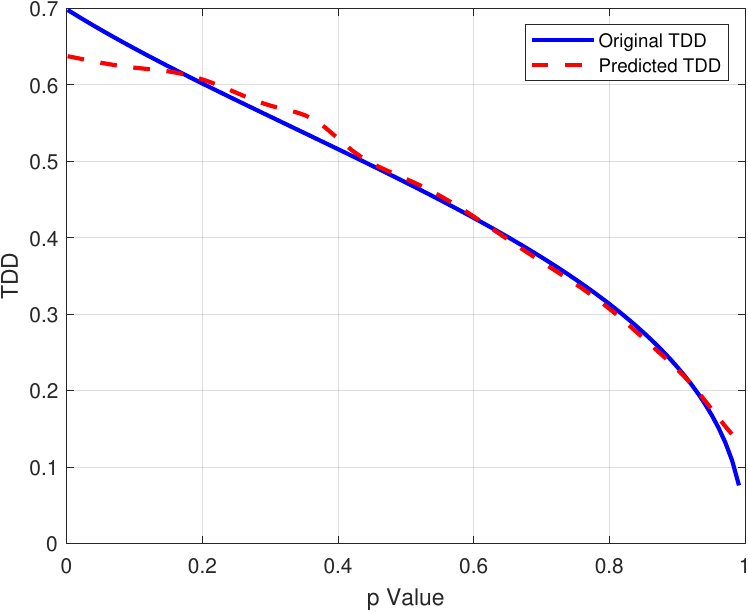} 
        \caption*{(b) Without WMR, \newline $\eta=1$.}
    \end{minipage}
    \begin{minipage}[b]{0.24\textwidth}
        \centering
        \includegraphics[width=\textwidth]{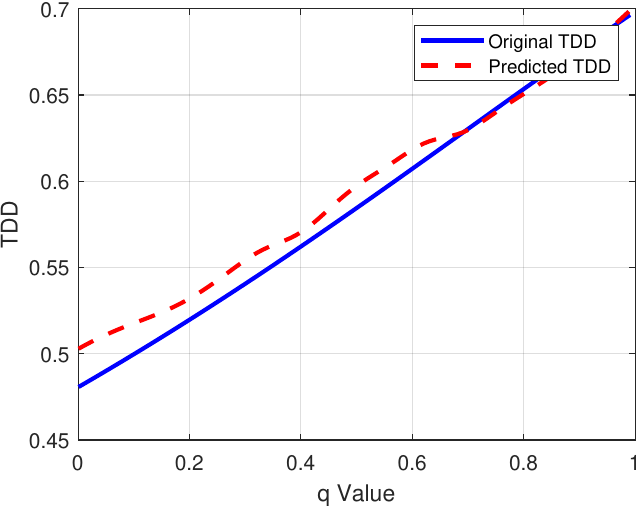} 
        \caption*{(c) WMR on both qubits, $\eta=0$.}
    \end{minipage}
    \begin{minipage}[b]{0.24\textwidth}
        \centering
        \includegraphics[width=\textwidth]{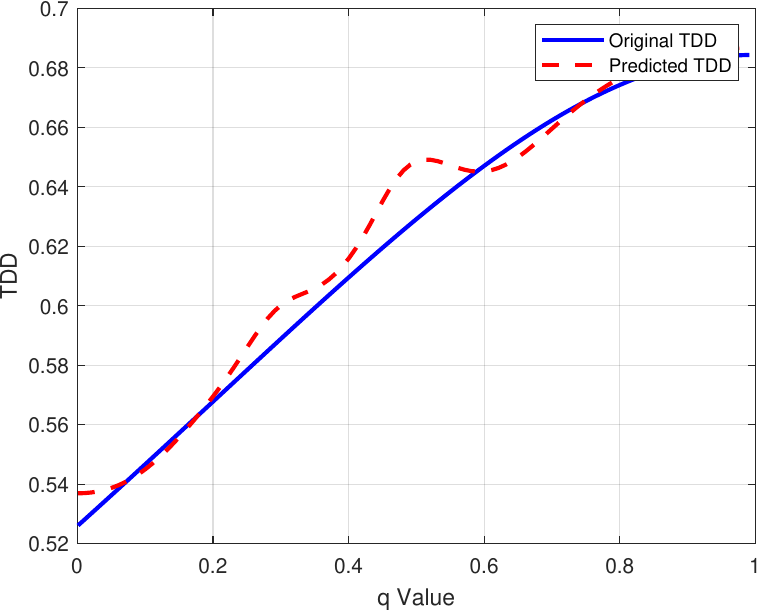} 
        \caption*{(d) WMR on both-qubits, $\eta=1$.}
    \end{minipage}
     \caption{\textit{Neural network predictions for TDD from JSD, Concurrence, teleportation Fidelity, EPR steering, and capacity of dense coding for Werner state ($r_b=0.8$). The dashed line shows the prediction against the actual value of TDD in a continuous line.}}
     \label{fig_TDDWState}
\end{figure}

\begin{figure}[H]
    \centering
    \begin{minipage}[b]{0.24\textwidth}
        \centering
        \includegraphics[width=\textwidth]{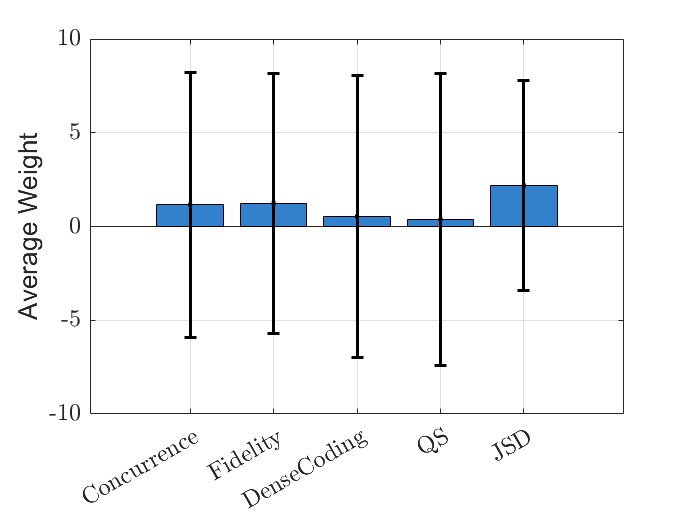} 
        \caption*{(a) Without WMR, \newline $\eta=0$.}
    \end{minipage}
    \begin{minipage}[b]{0.24\textwidth}
        \centering
        \includegraphics[width=\textwidth]{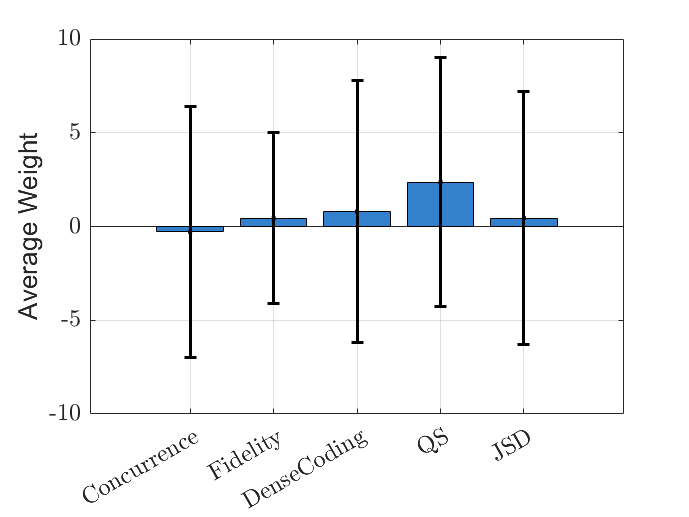} 
        \caption*{(b) Without WMR, \newline $\eta=1$.}
    \end{minipage}
    \begin{minipage}[b]{0.24\textwidth}
        \centering
        \includegraphics[width=\textwidth]{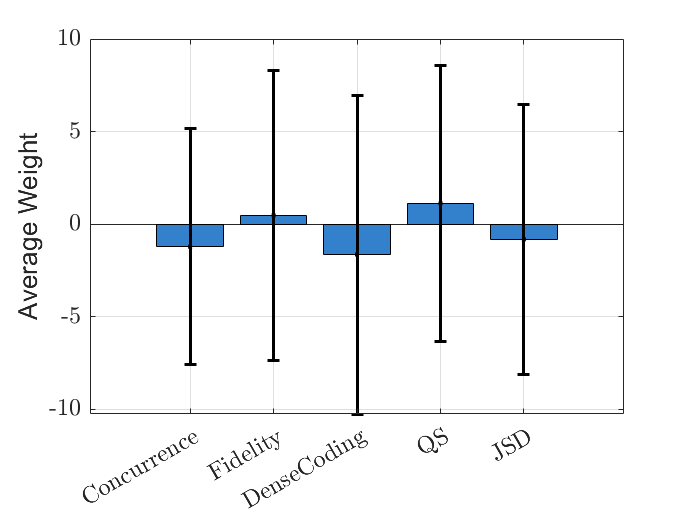} 
        \caption*{(c) WMR on both qubits, $\eta=0$.}
    \end{minipage}
    \begin{minipage}[b]{0.24\textwidth}
        \centering
        \includegraphics[width=\textwidth]{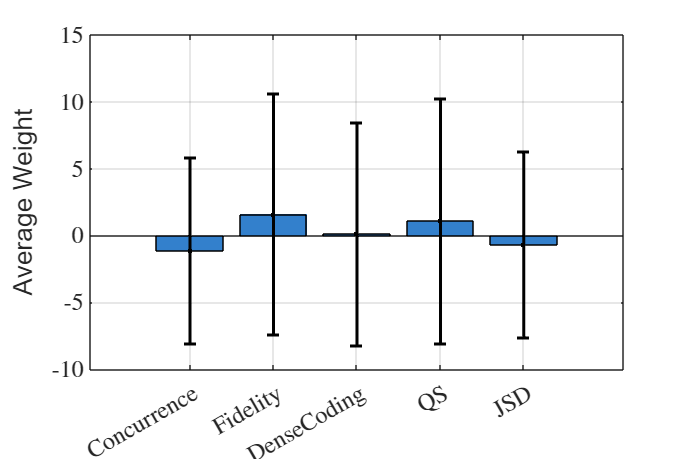} 
        \caption*{(d) WMR on both qubits, $\eta=1$.}
    \end{minipage}
     \caption{\textit{Average weights of neurons in the first layer with bars showing variations over different neurons in the prediction of TDD from JSD, Concurrence, teleportation Fidelity, EPR steering, and capacity of dense coding for Bell state as input.}}
     \label{fig_TDDWeights-Wstate}
\end{figure}

\section{Summary and Conclusion} 
\label{summary}

 We have studied the evolution in the quantum correlations of the Bell-, Werner-, and a maximally entangled mixed-state of two-qubits and their application to quantum information protocols in a correlated amplitude-damping channel. As correlations in the noise introduce greater structure to the noise, we expect that coherence properties will be better preserved under correlated noise than under uncorrelated noise of comparable strength. Based on this expectation, we found that entangled states are more robust in correlated noise channels than uncorrelated channels. In particular, the higher the correlation parameter, the better it preserves the entanglement, thus improving the performance of quantum information tasks such as dense coding, teleportation, and quantum steering. Normalized correlations typically show a pairwise trend: N[JSD] \& N[TDD], N[$C$] \& N[$\mathcal{F}$], and N[$\chi$] \& N[QS] in their quantitative behaviour under the CAD and AD channels.
 
 Applying the weak-measurement reversal protocols on one and both qubits in the decoherence process increases the correlations for all the states. As expected, the two-qubit WMR protocol performs better than the one-qubit WMR in protecting the coherence, entanglement, quantum steering, trace distance discord, dense coding capacity, and Fidelity; a perfect CAD channel helps in protecting most quantum correlations for all considered states better than a memoryless AD channel. However, we observe quantum steering for  MEMS as an exception where the AD channel performs better than CAD for WMR on two qubits. 
 
 \textcolor{black}{Notably, the loss of nonlocal correlations under noise follows a clear hierarchy: stronger quantum correlations vanish at lower noise strengths, while weaker quantum correlations persist even in strongly noisy environments. Moreover, the decay of quantum correlations under the influence of the CAD channel and their reversal under the WMR protocol of the considered states are found to be consistent with the hierarchy of quantum correlations \cite{Hierarchy_Paulson2016, Hierarchy_Paulson2021}. 
 As no single metric captures the full spectrum of quantumness in the presence of noise and the effect of corrective procedures, we have probed multiple measures (QS, $C$, TDD, JSD) that obey a strict ordering: there exist states with non-zero coherence and discord but without entanglement, and likewise states with non-zero entanglement but without steering. 
 Beyond this ladder, while teleportation fidelity $\mathcal{F}$ is a stronger measure than concurrence, yet it fails to be a strict entanglement monotone as previously observed for mixed states \cite{PhysRevA.66.022307, PhysRevA.81.054302}. Likewise, although QS and dense‐coding capacity $\chi$ are consistently stronger than all other measures, $\chi$ is not a strict monotone of QS \cite{e24081114}, exhibiting non‐monotonic revivals \cite{Haddadi_2024}. Importantly, channel memory further delays the sudden death of some correlations more than that of entanglement, and turns abrupt decays into asymptotic ones, underscoring why a comprehensive suite of correlation measures is essential to characterize and protect quantum resources under noise.} 
 
 Given the probabilistic nature of the reversal operation, the probability of success of the WMR protocol decreases with an increase in the strength of the weak-measurement operation. The present method could serve as an approach for quantum error mitigation in noisy intermediate-scale quantum computing and related applications. \textcolor{black}{WMR uses prior weak measurements and reversal operations to recover the quantum state probabilistically, making it effective for energy loss scenarios. The probability of success of the reversal operation depends on the strength of the weak measurement. On the other hand, the dynamic decoupling (DD) protocol applies rapid pulse sequences to average out environmental interactions, but it's more effective against dephasing than amplitude damping. WMR can outperform DD in non-Markovian or weak-damping regimes. WMR is better suited for amplitude damping, while DD is more general-purpose for phase noise.}

 Leveraging neural networks allowed us to explore complex relationships between quantum measures, highlighting the potential for machine learning to deepen the understanding of quantum correlations. These relationships also show the non-monotonicity between the measures. Future work could include investigating the effectiveness of machine-learning models for learning and predicting quantum measures in dynamic environments.

\section*{Appendix: A}
\label{Appendix_A}

Here, we define $p-1=\bar{p}'$, $q-1=\bar{q}'$ and $r-1=\bar{r}'$. \\
For Bell-state $\rho_B$, with WMR on one qubit, the evolution from Eq. (11) gives:\\
$ \rho_\text{WMR}[\rho_B]= \\
\resizebox{1\hsize}{!}{$
\left(\begin{array}{cccc}
\frac{{\left(\bar{r}'\right)}\,{\left(1+p^2 -p^2 \,q-\eta \,p+\eta \,p\,q-\eta \,p^2 +\eta \,p^2 \,q\right)}}{\sigma_1 } & 0 & 0 & -\frac{\sigma_2 }{r+{\left(\bar{q}'\right)}\,{\left(1-2\,p+2\,\eta \,p\right)}\,{\left(1-2\,p+p\,r\right)}-1}\\
0 & -\frac{p\,{\left(-1+\eta \right)}\,{\left(\bar{p}'\right)}\,{\left(\bar{q}'\right)}}{\sigma_1 } & 0 & 0\\
0 & 0 & \frac{p\,{\left(-1+\eta \right)}\,{\left(\bar{p}'\right)}\,{\left(\bar{q}'\right)}\,{\left(\bar{r}'\right)}}{\sigma_1 } & 0\\
-\frac{\sigma_2 }{\sigma_1 } & 0 & 0 & -\frac{{\left(\bar{p}'\right)}\,{\left(\bar{q}'\right)}\,{\left(\bar{p}+\eta \,p\right)}}{\sigma_1 }
\end{array}\right),
$}
\\
\mathrm{}\\
\textrm{where},\\
\mathrm{}\\
\;\;\sigma_1 =-2+r+q+4\,p-p\,r-4\,p\,q+p\,q\,r-4\,p^2 +2\,p^2 \,r+4\,p^2 \,q-2\,p^2 \,q\,r-2\,\eta \,p+2\,\eta \,p\,q+4\,\eta \,p^2 -2\,\eta \,p^2 \,r-4\,\eta \,p^2 \,q+2\,\eta \,p^2 \,q\,r,\;\;\sigma_2 =\sqrt{\bar{q}}\,\sqrt{\bar{r}}\,{\left(\eta \,p-p-\eta +\eta \,\sqrt{\bar{p}}+1\right)}.$
\\
\\
Corresponding Concurrence ($C(\rho_\text{WMR}[\rho_B])$), which is maximized to obtain optimal QMR strength $r$, is computed from Eq. (3) and given as \\
$
C(\rho_\text{WMR}[\rho_B])=\\
\resizebox{\textwidth}{!}{$
2\,\max \left(0,-\frac{\sqrt{\bar{r}}\,{\left(\eta \,\left|p\right|-\left|p\right|+p\,\left|p\right|+q\,\left|p\right|+\left|\eta \,p-p-\eta +\eta \,\sqrt{\bar{p}}+1\right|\,\sqrt{\bar{q}}-\eta \,p\,\left|p\right|-\eta \,q\,\left|p\right|-p\,q\,\left|p\right|+\eta \,p\,q\,\left|p\right|\right)}}{\sigma_1 },\frac{\sqrt{{\left(\bar{p}+\eta \,p\right)}\,\sigma_2 }\,\sqrt{\bar{p}}\,\sqrt{\bar{q}}\,\sqrt{\bar{r}}}{\sigma_1 }\right),
$}
\\
\mathrm{}\\
\textrm{where},\\
\mathrm{}\\
\;\;\sigma_1 =\left|\frac{{\left(\bar{r}'\right)}\,\sigma_2 }{-2+q}-\frac{{\left(-1+\eta \right)}\,{{\left(\bar{p}'\right)}}^2 \,{\left(\bar{q}'\right)}}{-2+q}-\frac{\eta \,{\left(\bar{p}'\right)}\,{\left(\bar{q}'\right)}}{-2+q}+\frac{\left|p\right|\,{\left(-1+\eta \right)}\,{\left(\bar{p}'\right)}\,{\left(\bar{q}'\right)}}{-2+q}-\frac{\left|p\right|\,{\left(-1+\eta \right)}\,{\left(\bar{p}'\right)}\,{\left(\bar{q}'\right)}\,{\left(\bar{r}'\right)}}{-2+q}\right|\,{\left(-2+q\right)},\\\sigma_2 =\eta \,\left|p\right|-\eta \,p^2 -p^2 \,q+p^2 +\eta \,p^2 \,q-\eta \,q\,\left|p\right|+1.$
\\
\\
For Bell-state $\rho_B$, with WMR on both qubits, the evolution from Eq. (11) gives:\\
$ \rho_\text{WMR}[\rho_B]= \\
\left(\begin{array}{cccc}
\frac{\sigma_3 }{\sigma_2 } & 0 & 0 & -\frac{{\left(\frac{2\,\sigma_4 \,{\left(-1+\eta \right)}\,{\left(\bar{p}'\right)}}{\sigma_5 }+\frac{2\,\eta \,\left|\bar{p}\right|\,\sigma_4 }{\sqrt{\bar{p}}\,\sigma_5 }\right)}\,{\left(\bar{r}'\right)}}{\sigma_2 }\\
0 & \sigma_1  & 0 & 0\\
0 & 0 & \sigma_1  & 0\\
-\frac{{\left(\bar{q}'\right)}\,{\left(\bar{r}'\right)}\,{\left(\eta \,p-p-\eta +\eta \,\sqrt{\bar{p}}+1\right)}}{\sigma_5 \,\sigma_2 } & 0 & 0 & -\frac{{{\left(\bar{q}'\right)}}^2 \,{\left(2\,\eta \,p-2\,p-\eta +\eta \,\left|\bar{p}\right|-\eta \,p^2 +p^2 +1\right)}}{\sigma_5 \,\sigma_2 }
\end{array}\right),\\
\mathrm{}\\
\textrm{where},\\
\mathrm{}\\
\;\;\sigma_1 =\frac{2\,p\,\left|\bar{p}\right|\,\left|\bar{r}\right|\,\sigma_4 \,{\left(-1+\eta \right)}\,{\left(\bar{q}'\right)}}{\sigma_5 \,\sigma_2 },\;\;\sigma_2 =\sigma_3 +\frac{2\,\sigma_4 \,{\left(-1+\eta \right)}\,{{\left(\bar{p}'\right)}}^2 \,{\left(\bar{q}'\right)}}{\sigma_5 }-\frac{2\,\eta \,\left|\bar{p}\right|\,\sigma_4 \,{\left(\bar{q}'\right)}}{\sigma_5 }+\frac{4\,p\,\left|\bar{p}\right|\,\left|\bar{r}\right|\,\sigma_4 \,{\left(-1+\eta \right)}\,{\left(\bar{q}'\right)}}{\sigma_5 },\;\;\sigma_3 ={\left(\frac{{\left(-1+\eta \right)}\,{\left(1+p^2 -2\,p^2 \,q+p^2 \,q^2 \right)}}{\sigma_5 }-\frac{\eta \,{\left(1+p-2\,p\,q+p\,q^2 \right)}}{\sigma_5 }\right)}\,{{\left(\bar{r}'\right)}}^2,\;\;\sigma_4 =-\frac{1}{2}+\frac{q}{2},\;\;\sigma_5 =2-2\,q+q^2.$
\\
\\
Corresponding Concurrence ($C(\rho_\text{WMR}[\rho_B])$), which is maximized to obtain optimal QMR strength $r$, is computed from Eq. (3) and given as \\
$
C(\rho_\text{WMR}[\rho_B])=\\
\resizebox{\textwidth}{!}{$2\,\max \left( 0,\frac{{\left(\bar{q}'\right)}\,{\left(\bar{r}'\right)}\,{\left(\left|\eta \,p-p-\eta +\eta \,\sqrt{\bar{p}}+1\right|-\left|p\right|+\eta \,\left|p\right|+p\,\left|p\right|+q\,\left|p\right|-\eta \,p\,\left|p\right|-\eta \,q\,\left|p\right|-p\,q\,\left|p\right|+\eta \,p\,q\,\left|p\right|\right)}}{\sigma_1 \,\sigma_4 },-\frac{\sqrt{-\sigma_2 \,{\left(\bar{p}+\eta \,p\right)}}\,\sqrt{\bar{p}}\,{\left(\bar{q}'\right)}\,{\left(\bar{r}'\right)}}{\sigma_1 \,\sqrt{\sigma_4 }}\right)$},\\
\mathrm{}\\
\textrm{where},\\
\mathrm{}\\
\;\;\sigma_1 =\left|-\sigma_2 \,{{\left(\bar{r}'\right)}}^2 -\frac{2\,\sigma_3 \,{\left(-1+\eta \right)}\,{{\left(\bar{p}'\right)}}^2 \,{\left(\bar{q}'\right)}}{\sigma_4 }-\frac{2\,\eta \,\sigma_3 \,{\left(\bar{p}'\right)}\,{\left(\bar{q}'\right)}}{\sigma_4 }-\frac{4\,\left|p\right|\,\sigma_3 \,{\left(-1+\eta \right)}\,{\left(\bar{p}'\right)}\,{\left(\bar{q}'\right)}\,{\left(\bar{r}'\right)}}{\sigma_4 }\right|,\;\;\sigma_2 =\frac{{\left(-1+\eta \right)}\,{\left(1+p^2 -2\,p^2 \,q+p^2 \,q^2 \right)}}{\sigma_4 }-\frac{\eta \,{\left(\left|p\right|+q^2 \,\left|p\right|-2\,q\,\left|p\right|+1\right)}}{\sigma_4 },\;\;\sigma_3 =-\frac{1}{2}+\frac{q}{2},\;\;\sigma_4 =2-2\,q+q^2.$

\section*{Appendix: B}
\label{Appendix_B}

For Werner state $\rho_W$, with WMR on one qubit, the evolution from Eq. (11) gives:
\\
\resizebox{\textwidth}{!}{$
\begin{aligned}
\rho^{11}_{QMR}[\rho_W] &= \frac{\left|\bar{r}\right|\,{\left(p+r_b-\eta \,p-p\,r_b+p\,\left|\bar{q}\right|+\sigma_{14} -\sigma_{13} +\sigma_{12} +\eta \,p\,r_b-p\,r_b\,\left|\bar{q}\right|+\sigma_{11} -\sigma_9 +1\right)}}{\sigma_5}, \\
\rho^{14}_{QMR}[\rho_W] &= \frac{2\,r_b\,\left|\bar{q}\right|\,\sqrt{\bar{r}}\,{\left(\sigma_1 -\sigma_4 -p\,\sqrt{\bar{p}}+\sqrt{\bar{p}}+\eta \,p\,\sqrt{\bar{p}}\right)}}{\sqrt{\bar{p}}\,\sqrt{\bar{q}}\,\sigma_5 }, \\
\rho^{22}_{QMR}[\rho_W] &= \frac{\left|\bar{q}\right|\,{\left(\eta +\left|\bar{p}\right|-\eta \,r_b-\sigma_1 +p\,\left|\bar{p}\right|-\sigma_3 -\eta \,p\,\left|\bar{p}\right|+\sigma_2 +p\,r_b\,\left|\bar{p}\right|-\eta \,p\,r_b\,\left|\bar{p}\right|\right)}}{\sigma_5}, \\
\rho^{33}_{QMR}[\rho_W] &= \frac{\left|\bar{r}\right|\,{\left(\eta +\left|\bar{p}\right|-\eta \,r_b-\sigma_1 -\sigma_3 +\sigma_2 +\sigma_{10} -\sigma_8 +\sigma_7 -\sigma_6 \right)}}{\sigma_5 }, \\
\rho^{41}_{QMR}[\rho_W] &= \frac{2\,r_b\,\left|\bar{r}\right|\,\sqrt{\bar{q}}\,{\left(\eta \,p-p-\eta +\sigma_4 +1\right)}}{\sqrt{\bar{r}}\,\sigma_5 }, \\
\rho^{44}_{QMR}[\rho_W] &= \frac{\left|\bar{q}\right|\,{\left(1+r_b\right)}\,{\left(2\,\eta \,p-2\,p-\eta +\sigma_1 -\eta \,p^2 +p^2 +1\right)}}{\sigma_5 }
\end{aligned}
$}
\\
$\textrm{where}\\
\;\;\sigma_1 =\eta \,\left|\bar{p}\right|,\;\;\sigma_2 =\eta \,r_b\,\left|\bar{p}\right|,\;\;\sigma_3 =r_b\,\left|\bar{p}\right|,\;\;\sigma_4 =\eta \,\sqrt{\bar{p}},\;\;\sigma_5 =\left|\bar{q}\right|+\left|\bar{r}\right|+\eta \,\left|\bar{r}\right|-2\,p\,\left|\bar{q}\right|+p\,\left|\bar{r}\right|+r_b\,\left|\bar{q}\right|+r_b\,\left|\bar{r}\right|+\left|\bar{p}\right|\,\left|\bar{q}\right|+\left|\bar{p}\right|\,\left|\bar{r}\right|+\sigma_{14} -\sigma_{13} +\sigma_{12} +p^2 \,\left|\bar{q}\right|\,\left|\bar{r}\right|+2\,\eta \,p\,\left|\bar{q}\right|-\eta \,p\,\left|\bar{r}\right|-2\,\eta \,r_b\,\left|\bar{q}\right|-\eta \,r_b\,\left|\bar{r}\right|-2\,p\,r_b\,\left|\bar{q}\right|-p\,r_b\,\left|\bar{r}\right|-\eta \,\left|\bar{p}\right|\,\left|\bar{r}\right|+\sigma_{10} +p\,\left|\bar{q}\right|\,\left|\bar{r}\right|-r_b\,\left|\bar{p}\right|\,\left|\bar{q}\right|-r_b\,\left|\bar{p}\right|\,\left|\bar{r}\right|+\sigma_{11} +\eta \,p\,r_b\,\left|\bar{r}\right|-\sigma_8 +2\,\eta \,r_b\,\left|\bar{p}\right|\,\left|\bar{q}\right|+\eta \,r_b\,\left|\bar{p}\right|\,\left|\bar{r}\right|+\sigma_7 -p\,r_b\,\left|\bar{q}\right|\,\left|\bar{r}\right|+p\,\left|\bar{p}\right|\,\left|\bar{q}\right|\,\left|\bar{r}\right|-\sigma_9 -\eta \,p^2 \,\left|\bar{q}\right|\,\left|\bar{r}\right|+p^2 \,r_b\,\left|\bar{q}\right|\,\left|\bar{r}\right|-\eta \,p^2 \,r_b\,\left|\bar{q}\right|\,\left|\bar{r}\right|-\sigma_6 +2\,\eta \,p\,r_b\,\left|\bar{q}\right|\,\left|\bar{r}\right|-\eta \,p\,\left|\bar{p}\right|\,\left|\bar{q}\right|\,\left|\bar{r}\right|+p\,r_b\,\left|\bar{p}\right|\,\left|\bar{q}\right|\,\left|\bar{r}\right|-\eta \,p\,r_b\,\left|\bar{p}\right|\,\left|\bar{q}\right|\,\left|\bar{r}\right|,\;\;\sigma_6 =\eta \,p\,r_b\,\left|\bar{p}\right|\,\left|\bar{q}\right|,\;\;\sigma_7 =p\,r_b\,\left|\bar{p}\right|\,\left|\bar{q}\right|,\;\;\sigma_8 =\eta \,p\,\left|\bar{p}\right|\,\left|\bar{q}\right|,\;\;\sigma_9 =\eta \,p^2 \,r_b\,\left|\bar{q}\right|,\;\;\sigma_{10}=p\,\left|\bar{p}\right|\,\left|\bar{q}\right|,\;\;\sigma_{11} =2\,\eta \,p\,r_b\,\left|\bar{q}\right|,\;\;\sigma_{12} =p^2 \,r_b\,\left|\bar{q}\right|,\;\;\sigma_{13} =\eta \,p^2 \,\left|\bar{q}\right|,\;\;\sigma_{14} =p^2 \,\left|\bar{q}\right|$.
\\
\\
Corresponding Concurrence ($C(\rho_\text{WMR}[\rho_W])$), which is maximized to obtain optimal QMR strength $r$, is computed from Eq. (3) and given as \\
$C(\rho_\text{WMR}[\rho_W])=
2\,\max \left( 0,\frac{{\left(2\,\left|r_b\,{\left(\eta -\eta \,\sqrt{\bar{p}}+\sqrt{\bar{p}}\right)}\right|\,\sqrt{\bar{p}}-\sqrt{\sigma_2}\right)}\,\sqrt{\bar{q}}\,\sqrt{\bar{r}}}{\sigma_1 },-\frac{\sqrt{\bar{p}}\,\sqrt{\bar{q}}\,\sqrt{\bar{r}}\,\sqrt{\sigma_3}}{\sigma_1 }\right),\\
\textrm{where},\\
\;\;\sigma_1 =\left|-4+2\,r+2\,q+2\,p\,r-2\,p\,q\,r-\eta \,p\,r+\eta \,p\,r\,r_b+\eta \,p\,q\,r-\eta \,p\,q\,r\,r_b\right|, \;\;\sigma_2 = -{\left(-1+r_b-2\,p\,r_b+p^2 +p^2 \,r_b+2\,\eta \,p\,r_b-\eta \,p^2 -\eta \,p^2 \,r_b\right)}\,\left(\right.\eta -p-\eta \,r_b+\eta \,{\left(\bar{p}'\right)}+r_b\,{\left(\bar{p}'\right)}-\eta \,r_b\,{\left(\bar{p}'\right)}+p\,{\left(\bar{p}'\right)}\,{\left(\bar{q}'\right)}-\eta \,p\,{\left(\bar{p}'\right)}\,{\left(\bar{q}'\right)}+p\,r_b\,{\left(\bar{p}'\right)}\,{\left(\bar{q}'\right)}-\eta \,p\,r_b\,{\left(\bar{p}'\right)}\,{\left(\bar{q}'\right)}+1\left.\right),\;\;\sigma_3={\left(1+r_b\right)}\,{\left(\bar{p}+\eta \,p\right)}\,\left(\right.p+r_b-\eta \,p-p\,r_b-p\,{\left(\bar{q}'\right)}-p^2 \,{\left(\bar{q}'\right)}+p\,r_b\,{\left(\bar{q}'\right)}+\eta \,p^2 \,{\left(\bar{q}'\right)}-p^2 \,r_b\,{\left(\bar{q}'\right)}+\eta \,p\,r_b-2\,\eta \,p\,r_b\,{\left(\bar{q}'\right)}+\eta \,p^2 \,r_b\,{\left(\bar{q}'\right)}+1\left.\right).$
\\
\\
 For Werner state $\rho_W$, with WMR on two-qubit, the evolution from Eq. (11) gives:
\\
$\rho_\text{WMR}[\rho_W]=\\
\left(\begin{array}{cccc}
\frac{\sigma_5 }{\sigma_4 } & 0 & 0 & -\frac{2\,r_b\,{\left(\bar{q}'\right)}\,{\left(\bar{r}'\right)}\,{\left(\sigma_2 -\sigma_1 -p\,\sqrt{\bar{p}}+\sqrt{\bar{p}}+\eta \,p\,\sqrt{\bar{p}}\right)}}{\sqrt{\bar{p}}\,\sigma_4 \,\sigma_8 }\\
0 & \sigma_3  & 0 & 0\\
0 & 0 & \sigma_3  & 0\\
-\frac{2\,r_b\,{\left(\bar{q}'\right)}\,{\left(\bar{r}'\right)}\,{\left(\eta \,p-p-\eta +\sigma_1 +1\right)}}{\sigma_4 \,\sigma_8 } & 0 & 0 & -\frac{{{\left(\bar{q}'\right)}}^2 \,{\left(1+r_b\right)}\,{\left(2\,\eta \,p-2\,p-\eta +\sigma_2 -\eta \,p^2 +p^2 +1\right)}}{\sigma_4 \,\sigma_8 }
\end{array}\right),
\mathrm{}\\
\textrm{where},
\mathrm{}\\
\;\;\sigma_1 =\eta \,\sqrt{\bar{p}},\;\;\sigma_2 =\eta \,\left|\bar{p}\right|,\;\;\sigma_3 =\frac{\left|\bar{r}\right|\,\sigma_6 }{\sigma_4 },\;\;\sigma_4 =2\,\left|\bar{r}\right|\,\sigma_6 +\sigma_5 +\frac{4\,\sigma_7 \,{\left(-1+\eta \right)}\,{{\left(\bar{p}'\right)}}^2 \,{{\left(\bar{q}'\right)}}^2 }{\sigma_8 }-\frac{4\,\eta \,\left|\bar{p}\right|\,\sigma_7 \,{{\left(\bar{q}'\right)}}^2 }{\sigma_8 },\;\;\sigma_5 =\left(\frac{{\left(-1+\eta \right)}\,{\left(r_b+2\,p\,\left|\bar{q}\right|-2\,p^2 \,q+p^2 \,r_b+p^2 +p^2 \,q^2 -2\,p^2 \,q\,r_b+p^2 \,q^2 \,r_b-2\,p\,r_b\,\left|\bar{q}\right|+1\right)}}{\sigma_8 }\right.\\\left.-\frac{\eta \,{\left(1+r_b\right)}\,{\left(1+p-2\,p\,q+p\,q^2 \right)}}{\sigma_8 }\right)\,{{\left(\bar{r}'\right)}}^2,\;\;\sigma_6 =\frac{\left|\bar{p}\right|\,{\left(-1+\eta \right)}\,{\left(p+\left|\bar{q}\right|-2\,p\,q+p\,r_b-r_b\,\left|\bar{q}\right|+p\,q^2 +p\,q^2 \,r_b-2\,p\,q\,r_b\right)}}{\sigma_8 }+\frac{4\,\eta \,\left|\bar{q}\right|\,{\left(-\frac{1}{4}+\frac{r_b}{4}\right)}}{\sigma_8 },\;\;\sigma_7 =\frac{1}{4}+\frac{r_b}{4},\;\;\sigma_8 =2\,r_b-2\,q+2\,\left|\bar{q}\right|-2\,q\,r_b-2\,r_b\,\left|\bar{q}\right|+q^2 \,r_b+q^2 +2.
\\ \\
$
Corresponding Concurrence ($C(\rho_\text{WMR}[\rho_W])$), which is maximized to obtain optimal QMR strength $r$, is computed from Eq. (3) and given as 
$\\
C(\rho_\text{WMR}[\rho_W])= 2\,\max \left(0,\frac{{\left(\bar{r}'\right)}\,{\left(2\,q\,\sigma_1 -2\,\sigma_1 +\left|\sigma_6 \right|\,\left|\sigma_4 \right|\right)}}{\left|\sigma_6 \right|\,\sigma_2 },-\frac{\sqrt{\bar{p}}\,{\left(\bar{q}'\right)}\,{\left(\bar{r}'\right)}\,\sqrt{-\frac{\sigma_3 \,{\left(1+r_b\right)}\,{\left(\bar{p}+\eta \,p\right)}}{\sigma_6 }}}{\sigma_2 }\right),\\
\mathrm{}
\textrm{where},
\mathrm{}\\
\;\;\sigma_1 =\left|r_b\,{\left(\eta \,p-p-\eta +\eta \,\sqrt{\bar{p}}+1\right)}\right|,\;\;\sigma_2 =\left|-\sigma_3 \,{{\left(\bar{r}'\right)}}^2 -2\,\sigma_4 \,{\left(\bar{r}'\right)}-\frac{4\,\sigma_5 \,{\left(-1+\eta \right)}\,{{\left(\bar{p}'\right)}}^2 \,{{\left(\bar{q}'\right)}}^2 }{\sigma_6 }\right.\\\left.-\frac{4\,\eta \,\sigma_5 \,{\left(\bar{p}'\right)}\,{{\left(\bar{q}'\right)}}^2 }{\sigma_6 }\right|,\;\;\sigma_3 =\frac{{\left(-1+\eta \right)}\,{\left(1+r_b+2\,p-2\,p\,r_b-2\,p\,q+2\,p\,q\,r_b+p^2 +p^2 \,r_b-2\,p^2 \,q-2\,p^2 \,q\,r_b+p^2 \,q^2 +p^2 \,q^2 \,r_b\right)}}{\sigma_6 }-\frac{\eta \,{\left(1+r_b\right)}\,{\left(1+p-2\,p\,q+p\,q^2 \right)}}{\sigma_6 },\;\;\sigma_4 =\frac{4\,\eta \,{\left(-\frac{1}{4}+\frac{r_b}{4}\right)}\,{\left(\bar{q}'\right)}}{\sigma_6 }-\frac{{\left(-1+\eta \right)}\,{\left(\bar{p}'\right)}\,{\left(\bar{q}'\right)}\,{\left(1-r_b+p+p\,r_b-p\,q-p\,q\,r_b\right)}}{\sigma_6 },\;\;\sigma_5 =\frac{1}{4}+\frac{r_b}{4},\;\;\sigma_6 =4-4\,q+q^2 +q^2 \,r_b.
$

\section*{Appendix: C}
\label{Appendix_C}

For MEMS $\rho_M$, with WMR on one qubit, the evolution from Eq. (11) gives:
\\
\resizebox{\textwidth}{!}{$\begin{aligned}
\rho^{11}_{QMR}[\rho_M] &= \left\lbrace \begin{array}{cl}
\frac{{\left(\bar{r}'\right)}\,{\left(2\,p-2\,p\,q+\gamma -2\,\gamma \,p+2\,\gamma \,p\,q+\gamma \,p^2 -\gamma \,p^2 \,q-2\,\eta \,p+2\,\eta \,p\,q+3\,\eta \,\gamma \,p-3\,\eta \,\gamma \,p\,q-\sigma_3 +\eta \,\gamma \,p^2 \,q\right)}}{\sigma_4 } & \sigma_6 \\
\frac{{\left(\bar{r}'\right)}\,{\left(1+p-p\,q+p^2 -p^2 \,q-\eta \,p^2 +\eta \,p^2 \,q\right)}}{\sigma_1 } & \;0\le \gamma < \frac{2}{3}
\end{array}\right., \\
\rho^{14}_{QMR}[\rho_M] &= \rho^{41}_{QMR}[\rho_M] = \left\lbrace \begin{array}{cl}
\frac{\gamma \,\sqrt{\bar{q}}\,\sqrt{\bar{r}}\,\sigma_7 }{\gamma -\gamma \,r+{\left(\bar{q}'\right)}\,{\left(-2+2\,p\,r+\gamma -\gamma \,p\,r-2\,\eta \,p\,r+\sigma_8 \right)}} & \sigma_6 \\
-\frac{3\,\gamma \,\sqrt{\bar{q}}\,\sqrt{\bar{r}}\,\sigma_7 }{-6+2\,r+4\,q+4\,p\,r-4\,p\,q\,r-2\,\eta \,p\,r+\sigma_9 } & \;0\le \gamma < \frac{2}{3}
\end{array}\right., \\
\rho^{22}_{QMR}[\rho_M] &= \left\lbrace \begin{array}{cl}
-\frac{{\left(\bar{q}'\right)}\,{\left(-2+2\,p+2\,\gamma -3\,\gamma \,p+\gamma \,p^2 -2\,\eta \,p+3\,\eta \,\gamma \,p-\sigma_3 \right)}}{\sigma_4 } & \sigma_6 \\
\frac{{\left(\bar{q}'\right)}\,{\left(\bar{p}^2 +\eta \,p^2 \right)}}{\sigma_1 } & \;0\le \gamma < \frac{2}{3}
\end{array}\right., 
\\
\rho^{33}_{QMR}[\rho_M] &= \left\lbrace \begin{array}{cl}
-\frac{\gamma \,p\,{\left(-1+\eta \right)}\,{\left(\bar{p}'\right)}\,{\left(\bar{q}'\right)}\,{\left(\bar{r}'\right)}}{\sigma_4 } & \sigma_6 \\
-\frac{3\,p\,{\left(-\frac{1}{3}+\frac{q}{3}\right)}\,{\left(-1+\eta \right)}\,{\left(\bar{p}'\right)}\,{\left(\bar{r}'\right)}}{\sigma_1 } & \;0\le \gamma < \frac{2}{3}
\end{array}\right., \\
\end{aligned}$}
{$\begin{aligned}
\rho^{44}_{QMR}[\rho_M] &= \left\lbrace \begin{array}{cl}
-\frac{\gamma \,{\left(\bar{p}'\right)}\,{\left(\bar{q}'\right)}\,\sigma_2 }{\sigma_4 } & \sigma_6 \\
-\frac{{\left(\bar{p}'\right)}\,{\left(\bar{q}'\right)}\,\sigma_2 }{\sigma_1 } & \;0\le \gamma < \frac{2}{3}
\end{array}\right.,
\end{aligned}$}
\\$
\textrm{where},\\
\;\;\sigma_1 =-3+r+2\,q+2\,p\,r-2\,p\,q\,r-\eta \,p\,r+\eta \,p\,q\,r,\;\;\sigma_2 =\bar{p}+\eta \,p,\;\;\sigma_3 =\eta \,\gamma \,p^2,\;\;\sigma_4 =-2+2\,q+2\,p\,r-2\,p\,q\,r+\gamma \,r-\gamma \,q-\gamma \,p\,r+\gamma \,p\,q\,r-2\,\eta \,p\,r+\sigma_9 +\sigma_8 -2\,\eta \,\gamma \,p\,q\,r,\;\;\sigma_5,\;\;\sigma_6 =\;\textrm{if}\;\;\frac{2}{3}\le \gamma \le 1,\;\;\sigma_7 =\eta \,p-p-\eta +\eta \,\sqrt{\bar{p}}+1,\;\;\sigma_8 =2\,\eta \,\gamma \,p\,r,\;\;\sigma_9 =2\,\eta \,p\,q\,r.
$
\\
\\
Corresponding Concurrence ($C(\rho_\text{WMR}[\rho_M])$), which is maximized to obtain optimal QMR strength $r$, is computed from Eq. (3) and given as \\
$C(\rho_\text{WMR}[\rho_M])=
2\,\max \left( 0,\frac{\sqrt{\gamma }\,\sqrt{\bar{r}}\,{\left(\sqrt{\gamma }\,\left|\sigma_3 \right|\,\sqrt{\bar{q}}-\sigma_6 \,\sqrt{1-\eta }\,\sqrt{\bar{p}}+q\,\sigma_6 \,\sqrt{1-\eta }\,\sqrt{\bar{p}}\right)}}{\sigma_1 },\right.\\\left.-\frac{\sqrt{\gamma\bar{p}\bar{q}\bar{r} }\,\sqrt{\sigma_5 \,{\left(2\,p-2\,p\,q+\gamma -2\,\gamma \,p+2\,\gamma \,p\,q+\gamma \,p^2 -\gamma \,p^2 \,q-2\,\eta \,p+2\,\eta \,p\,q+3\,\eta \,\gamma \,p-3\,\eta \,\gamma \,p\,q-\sigma_7 +\eta \,\gamma \,p^2 \,q\right)}}}{\sigma_1 }\right), \frac{2}{3}\le \gamma \le 1 \\
C(\rho_\text{WMR}[\rho_M])=2\,\max \left(0,\frac{\sqrt{\bar{r}}\,{\left(3\,\left|\gamma \,\sigma_3 \right|\,\sqrt{\bar{q}}-2\,\sqrt{1-\eta }\,\sqrt{\bar{p}}\,\sigma_4 +2\,q\,\sqrt{1-\eta }\,\sqrt{\bar{p}}\,\sigma_4 \right)}}{2\,\sigma_2 },\right.\\\left.-\frac{\sqrt{\bar{p}}\,\sqrt{\bar{q}}\,\sqrt{\bar{r}}\,\sqrt{\sigma_5 \,{\left(1+p-p\,q+p^2 -p^2 \,q-\eta \,p^2 +\eta \,p^2 \,q\right)}}}{\sigma_2}\right),\;0\le \gamma < \frac{2}{3}\\
\textrm{where}\\
\;\;\sigma_1 =\left|2-2\,q-2\,p\,r+2\,p\,q\,r-\gamma \,r+\gamma \,q+\gamma \,p\,r-\gamma \,p\,q\,r+2\,\eta \,p\,r-2\,\eta \,p\,q\,r-2\,\eta \,\gamma \,p\,r\right.\\\left.+2\,\eta \,\gamma \,p\,q\,r\right|,\;\;\sigma_2 =\left|-3+r+2\,q+2\,p\,r-2\,p\,q\,r-\eta \,p\,r+\eta \,p\,q\,r\right|,\;\;\sigma_3 =\eta \,p-p-\eta +\eta \,\sqrt{\bar{p}}+1,\;\;\sigma_4 =\sqrt{p-p^3 +\eta \,p^3 },\;\;\sigma_5 =\bar{p}+\eta \,p,\;\;\sigma_6 =\sqrt{-p\,{\left(-2+2\,p+2\,\gamma -3\,\gamma \,p+\gamma \,p^2 -2\,\eta \,p+3\,\eta \,\gamma \,p-\sigma_7 \right)}},\;\;\sigma_7 =\eta \,\gamma \,p^2.$
\\
\\

For MEMS $\rho_M$, with WMR on two-qubit, the evolution from Eq. (11) gives:
\\
$\begin{aligned}
\rho^{11}_{QMR}[\rho_M] &= \left\lbrace \begin{array}{cl}
\frac{\sigma_7 }{\sigma_2 } & \sigma_1 \\
\frac{\sigma_{10} }{\sigma_{16} \,\sigma_3 } & \;0\le \gamma < \frac{2}{3}
\end{array}\right., \; \rho^{14}_{QMR}[\rho_M] = \left\lbrace \begin{array}{cl}
-\frac{\gamma \,{\left(\bar{q}'\right)}\,{\left(\bar{r}'\right)}\,\sigma_4 }{\sqrt{\bar{p}}\,\sigma_{15} \,\sigma_2 } & \sigma_1 \\
\frac{3\,\gamma \,{\left(\bar{q}'\right)}\,{\left(\bar{r}'\right)}\,\sigma_4 }{2\,\sqrt{\bar{p}}\,\sigma_{16} \,\sigma_3 } & \;0\le \gamma < \frac{2}{3}
\end{array}\right., \\ \rho^{41}_{QMR}[\rho_M] &= \left\lbrace \begin{array}{cl}
-\frac{\gamma \,{\left(\bar{q}'\right)}\,{\left(\bar{r}'\right)}\,\sigma_6 }{\sigma_{15} \,\sigma_2 } & \sigma_1 \\
\frac{3\,\gamma \,{\left(\bar{q}'\right)}\,{\left(\bar{r}'\right)}\,\sigma_6 }{2\,\sigma_{16} \,\sigma_3 } & \;0\le \gamma < \frac{2}{3}
\end{array}\right.,\; \rho^{22}_{QMR}[\rho_M] = \left\lbrace \begin{array}{cl}
\frac{\sigma_8 }{\sigma_2 } & \sigma_1 \\
\frac{\sigma_{11} }{\sigma_3 } & \;0\le \gamma < \frac{2}{3}
\end{array}\right., 
\\
\rho^{33}_{QMR}[\rho_M] &= \left\lbrace \begin{array}{cl}
\frac{\sigma_9 }{\sigma_{15} \,\sigma_2 } & \sigma_1 \\
-\frac{\sigma_{12} }{\sigma_{16} \,\sigma_3 } & \;0\le \gamma < \frac{2}{3}
\end{array}\right.,\; \rho^{44}_{QMR}[\rho_M] =\left\lbrace \begin{array}{cl}
-\frac{\gamma \,{{\left(\bar{q}'\right)}}^2 \,\sigma_5 }{\sigma_{15} \,\sigma_2 } & \sigma_1 \\
\frac{{{\left(\bar{q}'\right)}}^2 \,\sigma_5 }{\sigma_{16} \,\sigma_3 } & \;0\le \gamma < \frac{2}{3}
\end{array}\right., \\
\end{aligned}$
\\
{$\textrm{where},
\mathrm{}\\
\;\;\sigma_1 =\;\textrm{if}\;\;\frac{2}{3}\le \gamma \le 1,\;\;\sigma_2 =\sigma_7 +\sigma_8 +\frac{\gamma \,{\left(-1+\eta \right)}\,{{\left(\bar{p}'\right)}}^2 \,{{\left(\bar{q}'\right)}}^2 }{\sigma_{15} }-\frac{\eta \,\gamma \,\left|\bar{p}\right|\,{{\left(\bar{q}'\right)}}^2 }{\sigma_{15} }+\frac{\sigma_9 }{\sigma_{15} },\;\;\sigma_3 =\sigma_{11} +\frac{\sigma_{10} }{\sigma_{16} }-\frac{3\,\sigma_{17} \,{\left(-1+\eta \right)}\,{{\left(\bar{p}'\right)}}^2 \,{\left(\bar{q}'\right)}}{\sigma_{16} }+\frac{3\,\eta \,\left|\bar{p}\right|\,\sigma_{17} \,{\left(\bar{q}'\right)}}{\sigma_{16} }-\frac{\sigma_{12} }{\sigma_{16} },\;\;\sigma_4 =\sigma_{13} -\sigma_{14} -p\,\sqrt{\bar{p}}+\sqrt{\bar{p}}+\eta \,p\,\sqrt{\bar{p}},\;\;\sigma_5 =2\,\eta \,p-2\,p-\eta +\sigma_{13} -\eta \,p^2 +p^2 +1,\;\;\sigma_6 =\eta \,p-p-\eta +\sigma_{14} +1,\;\;\sigma_7 ={\left(\frac{{\left(-1+\eta \right)}\,{\left(\gamma +2\,p\,\left|\bar{q}\right|+\gamma \,p^2 -2\,\gamma \,p^2 \,q+\gamma \,p^2 \,q^2 -2\,\gamma \,p\,\left|\bar{q}\right|\right)}}{\sigma_{15} }-\frac{\eta \,\gamma \,{\left(1+p-2\,p\,q+p\,q^2 \right)}}{\sigma_{15} }\right)}\,{{\left(\bar{r}'\right)}}^2,\;\;\sigma_8 =\left|\bar{r}\right|\,{\left(\frac{2\,\eta \,\left|\bar{q}\right|\,{\left(-1+\gamma \right)}}{\sigma_{15} }+\frac{\left|\bar{p}\right|\,{\left(-1+\eta \right)}\,{\left(\sigma_{19} +\gamma \,p-\sigma_{18} +\gamma \,p\,q^2 -2\,\gamma \,p\,q\right)}}{\sigma_{15} }\right)},\;\;\sigma_9 =\gamma \,p\,\left|\bar{p}\right|\,\left|\bar{r}\right|\,{\left(-1+\eta \right)}\,{{\left(\bar{q}'\right)}}^2,\\\sigma_{10} ={{\left(\bar{r}'\right)}}^2 \,\left(\eta \,p+p\,\left|\bar{q}\right|-\eta \,p^2 -2\,p^2 \,q+p^2 +p^2 \,q^2 +\eta \,p\,q^2 +2\,\eta \,p^2 \,q-\eta \,p^2 \,q^2 -2\,\eta \,p\,q\right.\\\left.-\eta \,p\,\left|\bar{q}\right|+1\right),\;\;\sigma_{11} =\left|\bar{r}\right|\,{\left(\frac{\eta \,\left|\bar{q}\right|}{\sigma_{16} }-\frac{\left|\bar{p}\right|\,{\left(-1+\eta \right)}\,{\left(p+\left|\bar{q}\right|-2\,p\,q+p\,q^2 \right)}}{\sigma_{16} }\right)},\;\;\sigma_{13} =\eta \,\left|\bar{p}\right|,\;\;\sigma_{14} =\eta \,\sqrt{\bar{p}},\;\;\sigma_{12} =3\,p\,\left|\bar{p}\right|\,\left|\bar{r}\right|\,\sigma_{17} \,{\left(-1+\eta \right)}\,{\left(\bar{q}'\right)},\;\;\sigma_{15} =2\,\gamma +\sigma_{19} -2\,\gamma \,q-\sigma_{18} +\gamma \,q^2,\;\;\sigma_{16} =\left|\bar{q}\right|-2\,q+q^2 +2,\;\;\sigma_{17} =-\frac{1}{3}+\frac{q}{3},\;\;\sigma_{18} =2\,\gamma \,\left|\bar{q}\right|,\;\;\sigma_{19} =2\,\left|\bar{q}\right|
$}.
\\
\\
Corresponding Concurrence ($C(\rho_\text{WMR}[\rho_M])$), which is maximized to obtain optimal QMR strength $r$, is computed from Eq. (3) and given as \\
$C(\rho_\text{WMR}[\rho_M])=
2\,\max \left(0,\frac{\sqrt{\gamma }\,{\left(\bar{q}'\right)}\,{\left(\bar{r}'\right)}\,{\left(\sqrt{\gamma }\,\left|\eta \,p-p-\eta +\sigma_6 +1\right|-\sqrt{\sigma_7}\,\sqrt{1-\eta }\,\sqrt{\bar{p}}\,\sqrt{\sigma_3 }\right)}}{\sigma_1 },-\frac{\sqrt{\gamma }\,\sqrt{\sigma_8}\,\sqrt{\bar{p}}\,{\left(\bar{q}'\right)}\,{\left(\bar{r}'\right)}}{\sigma_1 }\right), \\ \;\frac{2}{3}\le \gamma \le 1 \\
C(\rho_\text{WMR}[\rho_M])=2\,\max \left(0,\frac{\sqrt{\bar{p}}\,{\left(\bar{r}'\right)}\,{\left(3\,q\,\sigma_5 -3\,\sigma_5 +2\,\sqrt{p\,{\left(\bar{p}\,q-p^2 +p^2 \,q+\eta \,p\,q+\eta \,p^2 -\eta \,p^2 \,q\right)}}\,\sqrt{1-\eta }\,{{\left(\bar{q}\right)}}^{3/2} \right)}}{2\,\sigma_2 },\right.\\\left.-\frac{\sqrt{\bar{p}}\,{\left(\bar{q}'\right)}\,{\left(\bar{r}'\right)}\,\sqrt{\sigma_4 \,{\left(1+p-p\,q+p^2 -2\,p^2 \,q+p^2 \,q^2 -\eta \,p\,q+\eta \,p\,q^2 -\eta \,p^2 +2\,\eta \,p^2 \,q-\eta \,p^2 \,q^2 \right)}}}{\sigma_2 }\right), \;0\le \gamma < \frac{2}{3},\;\\
\textrm{where},\\$
\resizebox{\textwidth}{!}{$
\begin{aligned}
\sigma_1 &= \left| 2 - 2r - 2q + 2qr - 2pr + 2pr^2 + 2pqr - 2pqr^2 + \gamma r^2 - 2\gamma qr + \gamma q^2 - 2\gamma pr^2 \right. \\
&\quad + 2\gamma pqr + 2\gamma pqr^2 - 2\gamma pq^2r + \gamma p^2r^2 - 2\gamma p^2qr^2 + \gamma p^2q^2r^2 + 2\eta pr - 2\eta pr^2 - 2\eta pqr \\
&\quad + 2\eta pqr^2 - 2\eta\gamma pr + 3\eta\gamma pr^2 + 2\eta\gamma pqr - 4\eta\gamma pqr^2 + \eta\gamma pq^2r^2 - \eta\gamma p^2r^2 \\
&\quad \left. + 2\eta\gamma p^2qr^2 - \eta\gamma p^2q^2r^2 \right|, \\[1ex]
\sigma_2 &= \left| 3 - 3r + r^2 - 3q + qr + q^2 - 3pr + pr^2 + 5pqr - pqr^2 - 2pq^2r + p^2r^2 - 2p^2qr^2 \right. \\
&\quad \left. + p^2q^2r^2 + \eta pr - \eta pqr - \eta pqr^2 + \eta pq^2r^2 - \eta p^2r^2 + 2\eta p^2qr^2 - \eta p^2q^2r^2 \right|, \\[1ex]
\sigma_3 &= 2 - 2q + \gamma q^2, \quad \sigma_4 = \bar{p} + \eta p, \quad \sigma_5 = \left|\gamma\left(\eta - \sigma_6 + \sqrt{\bar{p}}\right)\right|, \quad \sigma_6 = \eta \sqrt{\bar{p}}, \\[1ex]
\sigma_7 &= p\left(\frac{2\eta(-1+\gamma)(\bar{q}')}{\sigma_3} + \frac{(1-\eta)(\bar{p}')(\bar{q}')(-2 + 2\gamma - \gamma p + \gamma pq)}{\sigma_3} \right), \\[1ex]
\sigma_8 &= \sigma_4 \left(2p - 2pq + \gamma - 2\gamma p + 2\gamma pq + \gamma p^2 - 2\gamma p^2q + \gamma p^2q^2 - 2\eta p + 2\eta pq \right. \\
&\quad \left. + 3\eta\gamma p - 4\eta\gamma pq + \eta\gamma pq^2 - \eta\gamma p^2 + 2\eta\gamma p^2q - \eta\gamma p^2q^2 \right).
\end{aligned}
$}

\section*{Acknowledgements}
RS acknowledges partial support from the Indian Science \& Engineering Research Board (SERB) grant CRG /2022 /008345.

 \footnotesize 
 \setlength{\bibsep}{0pt plus 0.3ex}  
  \bibliographystyle{unsrt}
 \bibliography{reference.bib}

\end{document}